\documentclass[12pt]{article}
\usepackage{amsbsy}
\usepackage{amsmath}
\usepackage{amssymb}
\usepackage{amsthm}
\usepackage{float}
\usepackage{bm}
\usepackage{mathrsfs}
\usepackage{enumerate}
\usepackage{multirow}
\usepackage{multicol}
\usepackage{array}
\usepackage{color}
\usepackage{epsfig}
\usepackage{graphicx}	
\usepackage{subfigure}
\usepackage{booktabs}
\usepackage{times}
\usepackage{algorithm}
\usepackage{algorithmic}
\usepackage{threeparttable}
\usepackage[title]{appendix}
\usepackage[T1]{fontenc}
\usepackage[utf8]{inputenc}
\usepackage{natbib}
\usepackage{hyperref}
\usepackage{chngcntr}

\newtheorem{theorem}{Theorem}

\newtheorem{remark}{Remark}

\newtheorem{assumption}{Assumption}

\def\bxi{\boldsymbol \xi}
\def\bnu{\boldsymbol \nu}
\def\bz{{\boldsymbol z}}
\def\bW{{\boldsymbol W}}
\def\bv{{\boldsymbol v}}
\def\bgamma{\boldsymbol{\gamma}}
\def\tbx{\tilde{{\boldsymbol x}}}
\def\bx{\boldsymbol x}

\def\hbbeta{\hat{\boldsymbol{\beta}}}
\def\bbeta{\boldsymbol{\beta}}

\def\bpsi{\boldsymbol{\psi}}

\newcommand{\blind}{1}

\date{}
\begin{document}

	\def\spacingset#1{\renewcommand{\baselinestretch}%
		{#1}\small\normalsize} \spacingset{1}

	
	\if1\blind
	{
		\title{\bf Change-plane Analysis in Functional Response Quantile Regression}
		\author{Xin Guan\\
			School of Statistics and Mathematics, Zhongnan University of Economics and Law\\
			and \\
			Yiyuan Li\\
			School of Statistics and Management, Shanghai University of Finance and Economics\\
			and \\
			Xu Liu\thanks{Liu's work was supported partially by the National Natural Science Foundation of China (12271329,72331005) and the Program for Innovative Research Team of SUFE, the Shanghai Research Center for Data Science and Decision Technology, and the Open Research Fund of Yunnan Key Laboratory of Statistical Modeling and Data Analysis, Yunnan University.}\hspace{.2cm}\\
			School of Statistics and Management, Shanghai University of Finance and Economics\\
			and \\
			Jinhong You\thanks{The research of You was supported in part by the National Natural Science Foundation of China (11971291).}\hspace{.2cm} \\
			School of Statistics and Management, Shanghai University of Finance and Economics 
		}
		\maketitle
	} \fi
	
	\if0\blind
	{
		\bigskip
		\bigskip
		\bigskip
		\begin{center}
			{\LARGE\bf Change-plane Analysis in Functional Response Quantile Regression}
		\end{center}
		\medskip
	} \fi
	
\bigskip
\begin{abstract}
Change-plane analysis is a pivotal tool for identifying subgroups within a heterogeneous population, yet it presents challenges when applied to functional data.
In this paper, we consider a change-plane model within the framework of functional response quantile regression, capable of identifying and testing subgroups in non-Gaussian functional responses with scalar predictors.
The proposed model naturally extends the change-plane method to account for the heterogeneity in functional data.
To detect the existence of subgroups, we develop a weighted average of the squared score test statistic, which has a closed form and thereby reduces the computational stress. 
An alternating direction method of multipliers algorithm is formulated to estimate the functional coefficients and the grouping parameters.
We establish the asymptotic theory for the estimates based on the reproducing kernel Hilbert space and derive the asymptotic distributions of the proposed test statistic under both null and alternative hypotheses.
Simulation studies are conducted to evaluate the performance of the proposed approach in subgroup identification and hypothesis test.
The proposed methods are also applied to two datasets, one from a study on China stocks and another from the COVID-19 pandemic.
\end{abstract}

\noindent%
\vfill

\newpage
\spacingset{1.5} 

\section{Introduction}
Functional data offers great opportunities in fields including neuromedical imaging, economics, and finance. 
The existing literature on functional data analysis is extensive, with a primary focus on two modeling types: functional linear regression models (FLMs) and functional response regression models (FRMs).
The FLMs explore the relationship between a scalar response and functional predictors \citep{li2007on,yao2017}, whereas FRMs describe the relationship between a functional response and scalar predictors \citep{li2007,zhu2012}. For a comprehensive review, refer to \cite{ramsay2005} and \cite{wang2016}.
Conventional functional regression models often assume homogeneity across individuals in the population. However, this assumption may not hold in practice and could result in model misspecification.

Subgroup analysis is crucial for detecting population heterogeneity and has been widely used in precision medicine and economic decision-making.
While most existing research on subgroup analysis focus on scalar data \citep{shen2015,ma2017}, recent advancements have developed statistical methods tailored for functional data.
For instance, in machine learning-based methods,  \cite{li2021} and \cite{wang2021} developed K-means algorithm to cluster heterogeneous functional data, but the statistical properties of this approach have not been thoroughly examined. 
In contrast, model-based methods such as the mixture model proposed by \cite{yao2011} and  \cite{jiang2021} required strict distributional assumptions, while \cite{zhang2022} and \cite{sun2024} introduced concave fusion penalty methods that allow intercepts and functional coefficients to vary across different subgroups.
However, some of these methods face challenges in terms of interpretability, while others lack a robust theoretical foundation. 

An alternative approach for subgroup identification is the change-plane model, also known as the threshold model.
The change-plane method divides subgroups using a hyperplane determined by a linear combination of variables, rather than relying on a single variable as in the classical change-point model \citep{bai1998}. 
The change-plane model has found broad application in empirical research. For instance, \cite{hansen2011} reviewed its use in economics, \cite{huang2020} developed a generalized change-plane model for personalized treatment within populations, \cite{deng2022} applied a change-plane model to a Cox proportional hazards model for survival data analysis, and \cite{wei2023} extended the method to longitudinal data. 
As an extension, \cite{su2019} and \cite{zhangyy2021} explored the change-plane model within the framework of robust regression.
One limitation of the change-plane method is its restriction to dividing the sample into only two subgroups. To address this,  \cite{lijialiang2018} extended the method to accommodate multiple thresholds, allowing for the partitioning of the population into several subgroups with varying covariate effects. 
Nonetheless, there has been limited research on applying change-plane analysis to functional data.

\begin{figure}[!htpb]
	\centering
	\subfigure[]{
		\includegraphics[height=5cm,width=6cm]{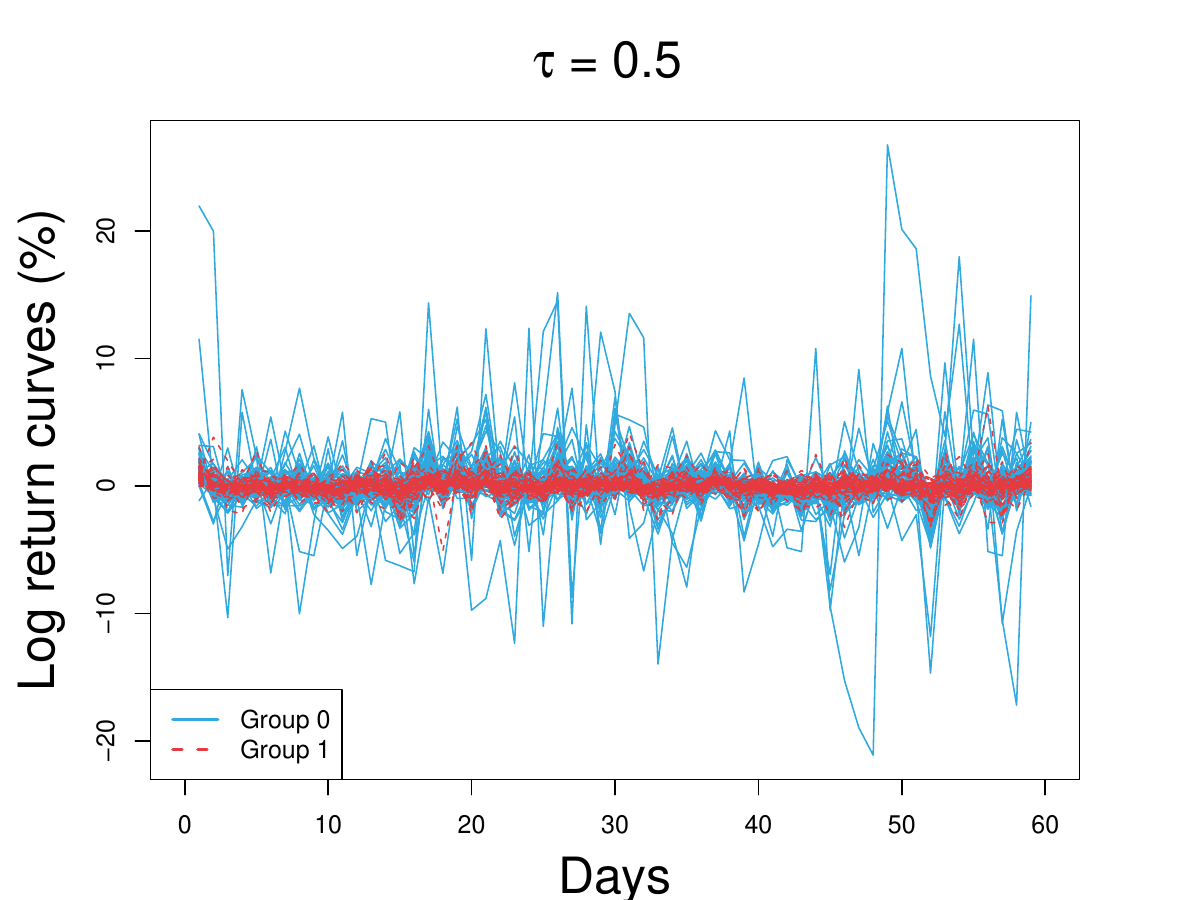}
	}
	\subfigure[]{
		\includegraphics[height=5cm,width=6cm]{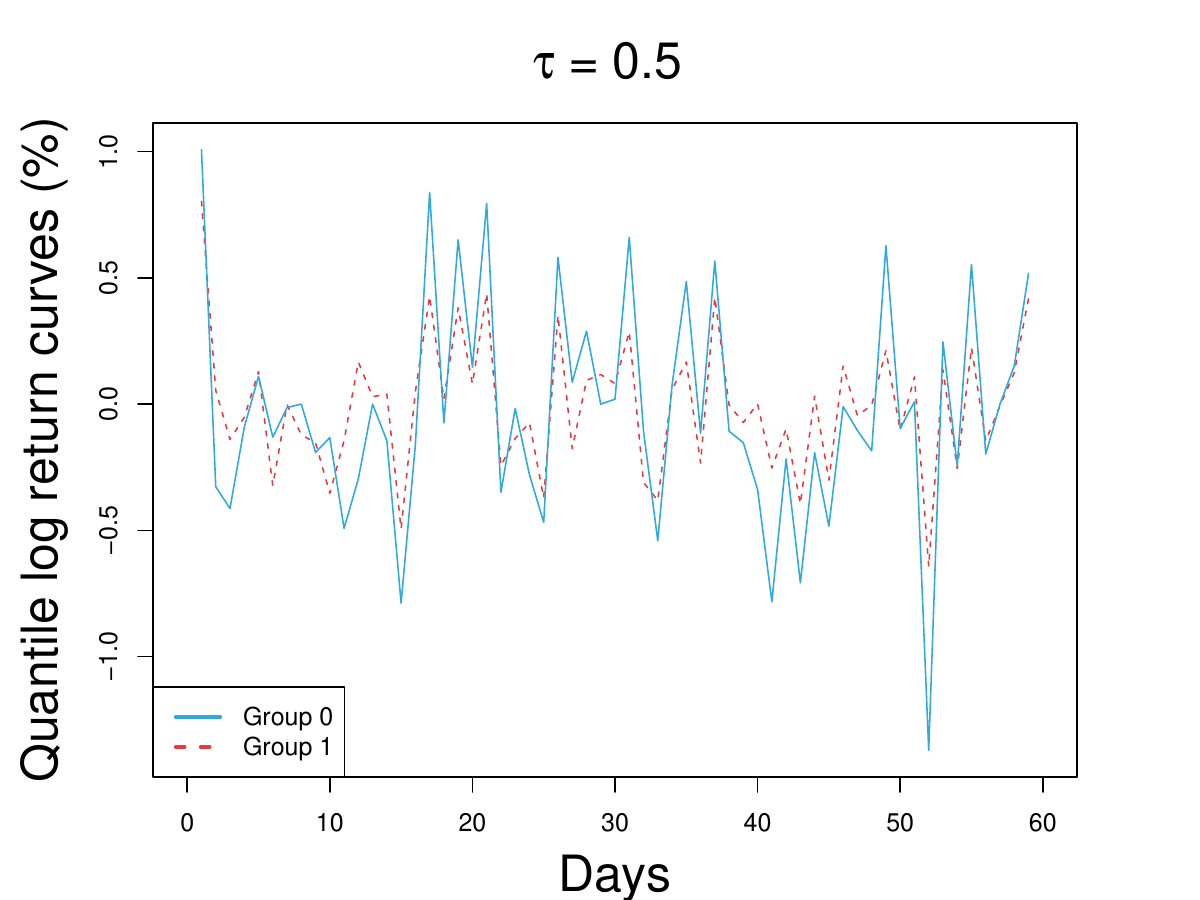}
	}
	\caption{\footnotesize{\small{(a) The log return curves of 182 stocks at the quantile level $\tau = 0.5$, where the dashed line represents Group 1 and the solid line represents Group 0; (b) The 0.5-th quantile of the log return curves for the two subgroups, where the dashed line represents Group 1 and the solid line represents Group 0.}}}
	\label{fig:stock_y50}
\end{figure}

This paper is motivated by the analysis of a stock dataset from the Shanghai Stock Exchange.
Figure \ref{fig:stock_y50}(a) shows the log return curves of 182 stocks, representing a typical example of functional data.
Through subgroup analysis, these 182 stocks are divided into two distinct groups, with Group 0 comprising 136 stocks and Group 1 consisting of 36 stocks; further details can be found in Section \ref{sec5}.
This indicates that the patterns of stock returns are heterogeneous across the identified subgroups.
Specifically, Figure \ref{fig:stock_y50}(a) reveals that
Group 1 is characterized by relatively stable log returns, while Group 0 exhibits heightened volatility. Additionally, Figure \ref{fig:stock_y50}(b) illustrates the distinct 0.5-th quantile log return curves for the two subgroups.
These figures suggest that  ignoring the heterogeneity within this stock dataset could lead to biased estimation and inference.
Moreover, the extreme observations in Figure \ref{fig:stock_y50}(a) indicate that Gaussian assumption may be inappropriate.

Driven by the challenges in modeling heterogeneous functional responses, we propose the following change-plane model within the framework of functional response quantile regression:
\begin{align}\label{eq1}
Q_{y_{i}(s)}(\tau | {\bx}_{i}, \widetilde{\bx}_i, {\bz}_{i}, s)={\bx}_{i}^{T} {\bm \beta}(s, \tau)+\widetilde{\bx}_i^{T} {\bm \theta}(s, \tau) I\left({\bz}_{i}^{T} {\bm \psi}({\tau}) \geq 0\right),
\end{align}
where $\tau \in (0,1)$ is a quantile of interest, $\{y_i(s), s \in \mathcal{S}\}$ denote the functional response process on a domain $ \mathcal{S}$, $Q_{y_{i}(s)}(\tau | \cdot)$ denotes the $\tau$-th conditional quantile of the functional response at a fixed location $s$ given predictors $({\bx}_{i}, \widetilde{\bx}_i, {\bz}_{i})$, and $I(\cdot)$ is the indicator function. The predictors ${\bx}_{i} \in \mathbb{R}^p$, $\widetilde{\bx}_i  \in \mathbb{R}^d (1 \leq d \leq p) $ is a subset of ${\bx}_i$, ${\bz}_{i} \in \mathbb{R}^{q+1}$ is the grouping variable, $\bm \beta(s, \tau)=(\beta_{1}(s, \tau), \cdots, \beta_{p}(s, \tau))^{T}$ and $\bm \theta(s, \tau)=(\theta_{1}(s, \tau), \cdots, \theta_{d}(s, \tau))^{T}$ are $p$-dimensional and $d$-dimensional vector of quantile regression coefficients, respectively, and ${\bm \psi} \in \mathbb{R}^{q+1}$ is the grouping parameter. 
Model (\ref{eq1}) establishes a hyperplane through linear combinations of the grouping variable $\bz_i$, allowing for different thresholds across subgroups.
It is evident that model ({\ref{eq1}}) requires an identification condition due to the presence of the indicator function. A classical normalization approach involves setting $\|\bm \psi(\tau)\|_2=1$, which imposes a constraint in the optimization algorithm. For computational convenience, however, we adopt an alternative identification condition as employed by \cite{seolinton2005} and \cite{zhangyy2021}. That is, we rewrite $\bz_i = (z_{1i}, \bz_{2i}^T)^T$ and $\bm \psi(\tau) = (1, \bm \gamma(\tau)^T)^T$ with $\bz_{2i}=(1, \widetilde{\bz}_{2i}^T)^T \in \mathbb{R}^q, \bgamma(\tau) \in \mathbb{R}^q$, 
so that ${\bz}_{i}^{T}\bm \psi(\tau) = z_{1i} + {\bz}_{2i}^{T}{\bgamma}({\tau})$.
Throughout this paper, we assume that the functional response $y_i(s)$ are observed at specified locations $\{s_j, 1 \leq j \leq m\}$ for all $i$, which was also considered in \cite{zhu2012} and  \cite{zhou2023}.

Model (\ref{eq1}) does not rely on a specific error distribution, making it flexible enough to describe how covariates influence the functional response at different quantile levels.
The optimization of model (\ref{eq1}) presents an infinite-dimensional challenge. To overcome this, various dimension reduction techniques can be employed, including functional principal component analysis (FPAC), spline, and reproducing kernel Hilbert space (RKHS) methods.
\cite{wahba1990} highlighted that the RKHS method yields an accurate estimation result, as opposed to the approximate solutions typically produced by FPAC or spline methods.
Although there is extensive literature on the RKHS method, most studies have focused on one-dimensional RKHS, see \cite{yuan2011} and \cite{shang2015}. However, the functional coefficients in model (\ref{eq1}) pertain to a multi-dimensional problem.

Inference on model (\ref{eq1}) is crucial, as neglecting to test for the existence of subgroups may result in false positive grouping outcomes.
\cite{songsui2017} proposed a supremum score test statistic to identify subgroups with an enhanced treatment effect, determining the supremum value through a grid search over the domain of the grouping parameter.
\cite{huang2020} developed a testing procedure based on the maximum of likelihood ratio statistics to assess the existence of subgroups associated with heterogeneity in disease risks. \cite{kang2022} developed an M-estimator to characterize two compound jump processes within the change-plane model.
While these methods exhibit sound statistical power, the process of identifying the extreme value is time-consuming and becomes increasingly challenging as the dimension of the parametric space increases.
Recently, \cite{liu2024} proposed a weighted average squared score test statistic with a closed-form solution, achieved by selecting an appropriate weight, thereby reducing the computational burden.
However, the aforementioned techniques are primarily focused on scalar data inference and cannot be directly applied to our case, as functional data inherently possesses infinite-dimensional characteristics.

The main contributions of this paper are summarized as follows. First, we extend change-plane analysis to model heterogeneous functional data, providing well-defined and easily interpretable subgroups.
The functional coefficients ${\bm \beta}(s,\tau)$ and ${\bm \theta}(s,\tau)$ in model (\ref{eq1}) are assumed to reside in an RKHS, leading to accurate estimation.
An alternating direction method of multipliers (ADMM) algorithm is developed to estimate the functional coefficients and grouping parameters. The accuracy rate of subgroup identification is calculated to assess the performance of the estimators.
Second, to enable statistical inference, we introduce a test statistic by taking the weighted average of the squared score test statistic (WAST) over the space of the grouping parameter.
The proposed WAST is a U-statistic with a closed-form expression, which significantly reduces computational complexity.
We also propose a weighted bootstrap procedure to approximate the critical value of the test statistic.
Third, we establish the asymptotic properties of the estimators for ${\bm \beta}(s,\tau)$ and ${\bm \theta}(s,\tau)$ within a vector RKHS. The asymptotic theory of the grouping parameter ${\bgamma}(\tau)$ is derived using a smoothing method for the indicator function, resulting in standard limiting distributions. The convergence rate for ${\bgamma}(\tau)$ is $\sqrt{h/n}$, where $h \rightarrow 0$ represents a bandwidth parameter.
We derive the asymptotic distributions of the proposed statistic under both null and alternative hypotheses. Additionally, we demonstrate the asymptotic consistency of the critical value obtained through the bootstrap method.

The paper is organized as follows. 
Section \ref{sec2} introduces a modified ADMM algorithm to estimate functional coefficients and grouping parameters, and establishes the asymptotic theory for estimators.
Section \ref{sec3} presents the development of a weighted average squared score test statistic for subgroup identification. 
Section \ref{sec4} provides simulation studies to evaluate the performance of the proposed method. Section \ref{sec5} demonstrates two applications using data from the China stock dataset. Section \ref{sec6} concludes the paper.
All technical proofs are relegated in the supplementary materials.

\section{Estimation procedure}\label{sec2}
Let $\mathcal{H}^{(r)}(\mathcal{S})$ be the $r$th order Sobolev space, which is abbreviate as $\mathcal{H}$ for simplicity:
\begin{align*}
\mathcal{H}^{(r)}(\mathcal{S}) = &\left\{f : \mathcal{S} \rightarrow \mathbb{R} | f^{(j)} \text{ is absolutely continuous for } j=1, \cdots r-1, f^{(r)} \in L_2(\mathcal{S})\right\},
\end{align*}
where $f^{(j)}$ is the $j$th derivative of $f(\cdot)$. Following \cite{cai2011aos} and  \cite{shang2015}, we assume that $r > 1/2$, ensuring that $\mathcal{H}$ is an RKHS.
Denote $K(\cdot, \cdot): \mathcal{S} \times \mathcal{S} \rightarrow \mathbb{R}$ as the reproducing kernel of $\mathcal{H}$.
The $(p+d)$-dimensional vector RKHS is defined as: $\mathcal{H}^{p+d} = \{(f_1, f_2, \cdots, f_{(p+d)}) : f_j \in \mathcal{H}, j = 1, \cdots, (p+d)\}.$
For details on the properties of multi-dimensional vector RKHS, refer to \cite{minh2016}.
Denote a $(p+d)$-dimensional functional coefficients ${\bm \alpha}(\cdot, \tau)^T = (\bm \beta(\cdot, \tau)^T, \bm \theta(\cdot, \tau)^T)$.  In this paper, we assume that each functional coefficient in model (\ref{eq1}) resides in $\mathcal{H}$, which implies that ${\bm \alpha} \in \mathcal{H}^{p+d}$.

Let $\{(y_i(s_j), {\bx}_i, \widetilde{\bx}_i, {\bz}_i), 1 \leq j \leq m\}_{i=1}^n$ be a sequence of independently identical distribution observations.
For a given $\tau$-th quantile level, the unknown functions and grouping parameters in the model (\ref{eq1}) can be obtained by solving the following optimization problem:
\begin{align}\label{eq2.1}
\min\limits_{(\bm \beta, \bm \theta, \bm \gamma)}  \frac{1}{nm}\sum_{i=1}^n \sum_{j=1}^m \rho_{\tau}\left(y_i(s_j)-{\bx}_{i}^{T}{\bm \beta}(s_j, \tau)-\widetilde{\bx}_i^{T} {\bm \theta}(s_j, \tau) I(z_{1i} + {\bz}_{2i}^{T}{\bgamma}({\tau}) \geq 0)\right) + \frac{\lambda}{2}J({\bm \alpha}, {\bm \alpha}),
\end{align}
where $\rho_{\tau}(u) = u\{\tau - I(u \leq 0)\}$ is the quantile check function, $J({\bm \alpha}, {\bm \alpha})$ is a roughness penalty that controls functional smoothness, and $\lambda$ is a tuning parameter. In this paper, we assume that $J({\bm \alpha}, {\bm \alpha}) = \sum_{k=1}^{p+d}J(\alpha_{k},\alpha_{k})$ with $J(\alpha_{k},\alpha_{k}) = \|\alpha_{k}\|_K^2$, where $\|\cdot \|_K$ is a semi-norm in $\mathcal{H}$. 
Refer to \cite{wahba1990} and  \cite{wangxiao2020} for similar assumptions regarding the penalty function.

As pointed out by \cite{yu2020}, directly estimating ${\bm \gamma}(\tau)$ in (\ref{eq2.1}) leads to a non-standard limiting distribution.
Alternatively, we use a smoothed function $G(\cdot)$ to approximate the indicator function, such as a cumulative distribution function, as employed in \cite{seolinton2005} and  \cite{zhangyy2021}.
That is, the smoothed estimator $\bm \eta({\tau})=(\bm \alpha(\cdot, \tau)^T, \bm \gamma({\tau})^T)^T$ can be defined by
\begin{align}\label{eq2.2}
\hat{\bm \eta}(\tau) = \mathop{\arg\min}\limits_{\bm \eta} \left\{\mathcal{L}_{nm}({\bm \eta};h) + \frac{\lambda}{2} J(\bm \alpha, \bm \alpha)\right\},
\end{align}
where 
$\mathcal{L}_{nm}({\bm \eta}; h) =({nm})^{-1}\sum_{i=1}^n \sum_{j=1}^m \rho_{\tau}\left\{y_i(s_j)-{\bx}_{i}^{T}{\bm \beta}(s_j, \tau)-\widetilde{\bx}_i^{T} {\bm \theta}(s_j, \tau) G_h(z_{1i} + {\bz}_{2i}^{T}{\bgamma}({\tau}))\right\}, 
$
$G_h(\cdot) = G(\cdot / h)$, and $h \rightarrow 0$ is a bandwidth parameter. Solving the optimization problem necessitates the use of an iterative algorithm.
For a given grouping parameter ${\bm \gamma}$, 
assume that $\tilde{\bm \beta}_{\bm \gamma}(\cdot, \tau) = (\tilde{\beta}_{k,\bm \gamma}(\cdot, \tau), 1 \leq k \leq p)^T$ and $\tilde{\bm \theta}_{\bm \gamma}(\cdot, \tau) = (\tilde{\theta}_{l,\bm \gamma}(\cdot, \tau), 1 \leq l \leq d)^T$ satisfy the following condition:
$$\left(\tilde{\bm \beta}_{\bm \gamma}, \tilde{\bm \theta}_{\bm \gamma}\right)=\mathop{\arg\min}\limits_{{\bm \beta}, {\bm \theta}} \mathcal{L}_{nm, \lambda}\left({\bm \eta}; h\right).$$
When the functional coefficients $(\tilde{\bm \beta}_{\bm \gamma}, \tilde{\bm \theta}_{\bm \gamma})$ are given, the grouping parameter can be estimated by 
$$\hat{\bm \gamma}(\tau) = \mathop{\arg\min}\limits_{\bm \gamma} \mathcal{L}_{nm}\left(\tilde{\bm \beta}_{\bm \gamma}, \tilde{\bm \theta}_{\bm \gamma}, \bm \gamma; h\right).$$
Therefore, the profiled function estimators can be given as $\hat{{\bm \beta}}_{\hat{\bm \gamma}}(\cdot,\tau) = \tilde{{\bm \beta}}_{\hat{\bm \gamma}}(\cdot,\tau)$, $\hat{{\bm \theta}}_{\hat{\bm \gamma}}(\cdot,\tau)  = \tilde{{\bm \theta}}_{\hat{\bm \gamma}}(\cdot,\tau)$.
Note that each component of $\bm \alpha$ resides in $\mathcal{H}$, according to the representation theorem \citep{wahba1990}, the profiled functional estimators have a finite form:
\begin{align}\label{eq2.3}
\tilde{\beta}_{k,\bm \gamma}(\cdot, \tau)= \xi_k + \sum_{j=1}^{m} b_{k j} K(\cdot, s_j), \quad  \tilde{\theta}_{l,\bm \gamma}(\cdot, \tau)= \nu_l + \sum_{j=1}^{m} c_{l j} K(\cdot, s_j),
\end{align}
where $\{\xi_k, \nu_l, b_{k j}, c_{l j}, 1 \leq k \leq p, 1 \leq l \leq d, 1\leq j \leq m\}$ are parameters to be estimated. 

The expression in (\ref{eq2.3}) demonstrates that the RKHS method simplifies the infinite-dimensional optimization problem into a finite-dimensional one through the representation theorem, requiring only the specification of the kernel function.
Some popular choices of $K(\cdot, \cdot)$ in practice are Gaussian kernel $K(s,t) = \exp(-\|s-t\|_2^2/(2\sigma^2))$, $q$th-degree polynomial kernel $K(s,t) = (\langle s,t \rangle + \sigma^2)^q$, Laplace kernel $K(s,t) = \exp(-\|s-t\|_1/\sigma)$, where $q$ and $\sigma$ are prespecified parameters.

\subsection{ADMM algorithm}\label{sec2.1}
In this subsection, we employ an ADMM algorithm to solve the optimization problem, following a similar approach to that used by \cite{Boyd2011} and  \cite{wangxiao2020}. 

Let $\boldsymbol{\varphi}^{T} = (\bxi^{T},\bnu^{T})$ with ${\bm \xi}^T=(\xi_1, \cdots, \xi_p)$ and ${\bm \nu}^T=(\nu_1, \cdots, \nu_d)$. 
Denote ${\bf d}^T=({\bf b}^T,{\bf c}^T)$, where ${\bf b}^T=({\bf b}_1^T, \cdots, {\bf b}_p^T)$ with ${\bf b}_k=(b_{kj}, 1 \leq j \leq m)^T$ for $1 \leq k \leq p$, and ${\bf c}^T=({\bf c}_1^T, \cdots, {\bf c}_d^T)$ with ${\bf c}_l=(c_{lj}, 1 \leq j \leq m)^T$ for $1 \leq l \leq d$. 
Denote $\bW_{i,\bm \gamma}^{T} = (\bx_i^{T}, \tbx_i^{T}G_h(z_{1i}+\boldsymbol{z}_{2i}^{T} \bgamma))\in \mathbb{R} ^{p+d}$.
Based on the expression in (\ref{eq2.3}), the regularized objective function in (\ref{eq2.2}) can be rewritten as follows:
\begin{align}
&\mathcal{L}_{nm,\lambda}(\boldsymbol{\varphi}, {\bf d},{\bm \gamma};h) = \frac{1}{nm} \sum_{i=1}^{n}\sum_{j=1}^{m} \rho_{\tau}\left\{y_i(s_j)- {\bx}_{i}^{T}{\bm \xi}-\widetilde{\bx}_{i}^{T}{\bm \nu}G_h(z_{1i} + {\bz}_{2i}^{T} {\bm \gamma})-\sum_{k=1}^p x_{ik}{\bf K}_{s_j}^T{\bf b}_k \right. \notag\\
&\qquad \qquad \qquad \qquad \qquad \qquad\left.-\sum_{l=1}^d \widetilde{x}_{il}{\bf K}_{s_j}^T{\bf c}_l G_h(z_{1i} + {\bz}_{2i}^{T} {\bm \gamma})\right\} + \frac{\lambda}{2} {\bf d}^T \Omega {\bf d},
\label{eq2.4}
\end{align}
where ${\bf K}=({\bf K}_{s_1}, \cdots,{\bf K}_{s_m})$ with ${\bf K}_{s} = (K(s,s_1),\cdots,K(s,s_m))^T$,  $\Omega=\left(\begin{array}{ll}{\bf I}_p & \\ & {\bf I}_d \end{array}\right) \otimes {\bf K}$, $\otimes$ is the Kronecker product, and ${\bf I}_p$ is a $p \times p$ identity matrix.

Equivalently, the optimization problem in (\ref{eq2.4}) can be expressed in the following form:
\begin{align*}
\min\limits_{(\boldsymbol{\varphi},{\bf d}, {\bm \gamma})} \quad &\frac{1}{nm} \sum_{i=1}^{n}\sum_{j=1}^{m} \rho_{\tau}\left(y_i(s_j) - u_{ij}\right) + \frac{\lambda}{2} {\bf d}^T \Omega {\bf d}, \\
\text{s.t.} \quad & u_{ij} = {\bx}_{i}^{T}{\bm \xi}+\widetilde{\bx}_{i}^{T}{\bm \nu}G_h(z_{1i} + {\bz}_{2i}^{T} {\bm \gamma})+ \sum_{k=1}^p x_{ik}{\bf K}_{s_j}^T{\bf b}_k + \sum_{l=1}^d \widetilde{x}_{il}{\bf K}_{s_j}^T{\bf c}_l G_h(z_{1i} + {\bz}_{2i}^{T} {\bm \gamma}), \\
& \quad  \quad  i=1,\cdots,n, j=1,\cdots,m.
\end{align*}
Using the augmented Lagrangian method, the estimates of parameters $(\boldsymbol{\varphi},{\bf d}, {\bm \gamma})$ can be obtained by minimizing the following objective function:
\begin{align}\label{eq:lag}
& L_{\sigma}({\boldsymbol u}, {\bm \zeta}, \boldsymbol{\varphi},{\bf d}, {\bm \gamma})  =  \frac{1}{nm} \sum_{i=1}^{n}\sum_{j=1}^{m} \rho_{\tau}\left(y_i(s_j) - u_{ij}\right) + \frac{\lambda}{2} {\bf d}^T \Omega {\bf d} \notag \\
& \quad  + \frac{1}{nm} \sum_{i=1}^{n}\sum_{j=1}^{m} \zeta_{ij}\left(u_{ij}-\psi_{ij}(\boldsymbol{\varphi  }, {\bf d},{\bm \gamma})\right) + \frac{\kappa}{2nm}\sum_{i=1}^{n}\sum_{j=1}^{m}\left(u_{ij}-\psi_{ij}(\boldsymbol{\varphi}, {\bf d},{\bm \gamma})\right)^2,
\end{align}
where ${\boldsymbol u} = \{u_{ij}, 1\leq i \leq n, 1\leq j \leq m\}$, ${\bm \zeta} = \{\zeta_{ij}, 1\leq i \leq n, 1\leq j \leq m\}$ are lagrange multipliers, 
$
\psi_{ij}(\boldsymbol{\varphi}, {\bf d},{\bm \gamma})={\bf x}_{i}^{T}{\bm \xi} + \widetilde{\bx}_{i}^{T}{\bm \nu}G_h(z_{1i} + {\bz}_{2i}^{T} {\bm \gamma}) + \sum_{k=1}^p x_{ik}{\bf K}_{s_j}^T{\bf b}_k + \sum_{l=1}^d \widetilde{x}_{il}{\bf K}_{s_j}^T{\bf c}_l G_h(z_{1i} + {\bz}_{2i}^{T} {\bm \gamma}),
$
and $\kappa$ is a penalty parameter. 
To obtain the estimate of $(\boldsymbol{\varphi},{\bf d}, {\bm \gamma})$, we develop a ADMM algorithm below, given the estimated parameters in $t$-th iteration.

The parameters are updated by
\begin{align}
u_{ij}^{(t+1)} :=& \underset{u_{ij}}{\rm argmin}\left(\rho_{\tau}\left(y_i(s_j)-u_{ij}\right)+\frac{\kappa}{2}\left(u_{ij}-
\psi_{ij}\left(\boldsymbol{\varphi}^{(t)},{\bf d}^{(t)},\bgamma^{(t)}\right)+\tilde{\zeta}_{ij}^{(t)}\right)^2\right), \label{eq:u}\\
\left(\boldsymbol{\varphi }^{(t+1)},{\bf d}^{(t+1)}\right) :=& \underset{(\boldsymbol{\varphi},{\bf d})}{\rm argmin}\frac{\kappa}{2}\sum_{i=1}^{n}\sum_{j=1}^{m}\left(u_{ij}^{(t+1)}-\psi_{ij}\left(\boldsymbol{\varphi},{\bf d},\bgamma^{(t)}\right)+\tilde{\zeta} _{ij}^{(t)}\right)^2 +\frac{\lambda}{2} {\bf d}^{T} \Omega {\bf d}, \notag \\
\bgamma^{(t+1)} :=& \underset{\bgamma}{\rm argmin}\frac{\kappa}{2}\sum_{i=1}^{n}\sum_{j=1}^{m}\left(u_{ij}^{(t+1)}-
\psi_{ij}\left(\boldsymbol{\varphi}^{(t+1)},{\bf d}^{(t+1)} ,\bgamma\right)+\tilde{\zeta} _{ij}^{(t)}\right)^2, \label{eq:gamma}\\
\tilde{\zeta}_{ij}^{(t+1)} :=& \tilde{\zeta} _{ij}^{(t)}+u_{ij}^{(t+1)}-\psi_{ij}\left(\boldsymbol{\varphi  }^{(t+1)},{\bf d}^{(t+1)},\bgamma^{(t+1)}\right), \label{eq:zeta}
\end{align} 
where $\tilde{\zeta}_{ij}^{(t)}=(1/\kappa)\zeta_{ij}^{(t)}$ is the scaled dual variable. 
Specifically, the optimal $u_{ij}$ can be explicitly solved using a proximal operator $S_{\tau,\sigma,v}(\cdot): \mathbb{R} \rightarrow \mathbb{R}$ defined by
$S_{\tau,\sigma,v}(u) ={\rm argmin}_u(\rho_{\tau}(u)+\frac{\sigma}{2}(u-v)^2),$
and the update of $u_{ij}$ can be written as
$u_{ij}^{(t+1)} = y_i(s_j)-S_{\tau,\sigma,y_i(s_j)-\psi_{ij}^{(t)}+\tilde{\zeta} _{ij}^{(t)}}\left(y_i(s_j)-u_{ij}\right).$
The optimal estimates of parameters $(\boldsymbol{\varphi}, {\bf d})$ can be obtained using the least squares method as follows:
\begin{equation}\label{eq:b}
\begin{aligned}
\boldsymbol{\varphi }^{(t+1)}=&\left(\sum_{i=1}^{n}\left\{\bW_{i,\bm \gamma}^{(t)}\right\}^{\otimes 2}\right)^{-1}\left(\sum_{i=1}^{n}\sum_{j=1}^{m}\bW_{i,\bm \gamma}^{(t)}\left(u_{ij}^{(t+1)}+\tilde{\zeta} _{ij}^{(t)}-\left\{\bW_{i,\bm \gamma}^{(t)}\right\}^{T} \otimes {\bf K}_{s_j}^{T}{\bf d}^{(t)} \right)\right),\\
{\bf d}^{(t+1)}=&\left(\sum_{i=1}^{n}\sum_{j=1}^{m}\left\{\bW_{i,\bm \gamma}^{(t)}\right\}^{\otimes 2} \otimes {\bf K}_{s_j}^{\otimes 2}+\frac{nm\lambda}{\kappa} \Omega \right)^{-1}
\Biggl\{\sum_{i=1}^{n}\sum_{j=1}^{m}\left(\bW_{i,\bm \gamma}^{(t)}\otimes {\bf K}_{s_j}\right)\\
& \times \left(u_{ij}^{(t+1)}-\left\{\bW_{i,\bm \gamma}^{(t)}\right\}^{T} \boldsymbol{\varphi}^{(t+1)} +\tilde{\zeta} _{ij}^{(t)}\right)  \Biggr\},
\end{aligned}  
\end{equation}    
where ${\bm v}^{\otimes 2} =\bm v {\bm v}^T$ for any vector $\bm v$.

We apply the stopping criterion recommended by \cite{Boyd2011} to determine when to stop the ADMM algorithm.
Calculate the primal residual $\chi_{ij}$ and dual residual $\phi_{ij}$ by
\begin{align}\label{eq:stop}
\chi_{ij}^{(t+1)} & = u_{ij}^{(t+1)}-\bW_{i,\bm \gamma}^{T} \boldsymbol{\varphi }^{(t+1)}-\bW_{i,\bm \gamma}^{T}\otimes {\bf K}_{s_j}^{T}{\bf d}^{(t+1)},\\
\phi_{ij}^{(t+1)} & =\kappa\left(\bW_{i,\bm \gamma}^{T}\otimes {\bf K}_{s_j}^{T}\right)\left({\bf d}^{(t+1)}-{\bf d}^{(t)}\right),  \notag 
\end{align}
respectively. Let $\chi^{(t+1)} = \{\chi_{ij}^{(t+1)}, 1\leq i \leq n, 1\leq j \leq m\}$ and $\phi^{(t+1)} = \{\phi_{ij}^{(t+1)}, 1\leq i \leq n, 1\leq j \leq m\}$. The iteration algorithm can be stopped when $\|\chi^{(t+1)}\|_2$ and $\|\phi^{(t+1)}\|_2$ are small than a pre-specified tolerance level.

The choice of regularization parameters is crucital in this optimization problem. In the simulation studies of Section \ref{sec5}, we determine the optimal value of $\lambda$ by minimizing the mean integrated squared error of the estimated parameters, while the Lagrange parameter $\kappa$ is fixed for computational convenience. Algorithm \ref{alg:description} provides a summary of the modified ADMM algorithm.
\vspace{-1cm}
\begin{algorithm}[!htbp]\footnotesize
	\caption{Modified ADMM Algorithm.}
	\label{alg:description}
	\begin{algorithmic}[1]
		\STATE \textbf{Input:} Datasets $ \left\{(y_i(s_j),\bx_i,\tbx_i,\bz_i ), j = 1,\cdots, m \right\}_{i=1}^n$, $\tau$-th quantile level, smoothing parameter $\kappa$. 
		\STATE Step {\bf1}: Give initial values of parameters  $(\boldsymbol{\varphi}^{(0)}, {\bf d}^{(0)},\bgamma^{(0)})$ and $(\boldsymbol{u}^{(0)}, {\bm \zeta}^{(0)})$.
		\STATE Step {\bf2}: Conduct an iteration algorithm to obtain estimates in objective function (\ref{eq:lag}). 
		\WHILE{$\Vert \chi \Vert_2\ge 10^{-3}$ and $\Vert \phi \Vert_2\ge 10^{-3}$}
		\STATE  Step 2.1: Update $\boldsymbol{u}^{(t+1)}$ by (\ref{eq:u}).
		\STATE  Step 2.2: Update $(\boldsymbol{\varphi }^{(t+1)}, {\bf d}^{(t+1)}) $ by (\ref{eq:b}).
		\STATE  Step 2.3: Update $ \bgamma^{(t+1)} $ by (\ref{eq:gamma}).
		\STATE Step 2.4: Update $\boldsymbol{\tilde{\zeta} }^{(t+1)}$ by (\ref{eq:zeta}).
		\STATE Step 2.5: Calculate $\Vert  \chi \Vert_2$ and $\Vert \phi \Vert_2$ by (\ref{eq:stop}).
		\ENDWHILE
		\STATE \textbf{Output:} The estimated parameters $(\hat{\varphi}, \hat{\bf d}, \hat{\bgamma}) = (\boldsymbol{\varphi }^{(t+1)},  {\bf d}^{(t+1)},\bgamma^{(t+1)})$.
		\label{alg:admm}
	\end{algorithmic}
\end{algorithm}

\subsection{Properities of Subgroup Identification}
Let ${\bv} = (\bx, \widetilde{\bx}, \bz)$ denote all predictors, ${e}_{i}(s, \tau) = y_{i}(s) - {\bx}_{i}^{T}{\bm \beta}_{0}(s, \tau) - \widetilde{\bx}_i^T{\bm \theta}_{0}(s, \tau) I({z_{1i} + {\bz}_{2i}^{T} {\bm \gamma}_0(\tau)} \geq 0)$, where ${\bm \alpha}_{0}(s, \tau)^T = ({\bm \beta}_{0}(s, \tau)^T, {\bm \theta}_{0}(s, \tau)^T)$ and ${\bm \gamma}_{0}(\tau)$ are the true coefficients given $\tau$-th quantile level. Denote ${\bm \eta}_{0}({\tau})^T = ({\bm \alpha}_{0}(s,\tau)^T, {\bm \gamma}_0({\tau})^T)$ and $\bW^{*T} = (\bx^{T}, {\tbx}^{T}I({z_{1i} + {\bz}_{2i}^{T} {\bm \gamma}_{0}(\tau)} \geq 0))$. For each $s \in \mathcal{S}$ and a given quantile level $\tau$, denote $F_{e}(a, s | {\bm v})$ and $f_{e}(a, s | {\bm v})$ as the conditional distribution and density functions of ${e}(s, \tau)$, respectively. Define $F_e(a_1, a_2, s, t | {\bm v}) = P(e(s, \tau) < a_1, e(t, \tau) < a_2 | {\bm v})$ as the bivariate cumulative distribution of individual effects $e(s, \tau)$ and $e(t, \tau)$. Let $f_e(a_1,a_2,s,t | {\bm v}) = {\partial^2 F_e(a_1,a_2,s,t | {\bm v})}/{\partial a_1 \partial a_2}$. 

Denote $\mathcal{D}$ as the Fr\'{e}chet derivative operator.
Let $S_{nm}^{\bm \alpha}({\bm \eta};h)$ and $S_{nm,{\lambda}}^{\bm \alpha}({\bm \eta};h)$ be the first Fr\'{e}chet derivative of $\mathcal{L}_{nm}({\bm \eta}; h)$ and $\mathcal{L}_{nm,{\lambda}}({\bm \eta}; h)$ with respect to ${\bm \alpha}$, respectively. 
For any ${\bm \alpha}_1, {\bm \alpha}_2 \in \mathcal{H}^{p+d}$, we have
\begin{align*}
&S_{nm}^{\bm \alpha}({\bm \eta};h) {\bm \alpha}_1 =-\frac{1}{nm} \sum_{i=1}^{n}\sum_{j=1}^m \left\{\tau -I\left(y_{i}(s_j) \leq {\bW}_{i,\bm \gamma}^{T} {\bm \alpha}(s_j, \tau)\right)\right\} {\bW}_{i,\bm \gamma}^{T} {\bm \alpha}_1(s_j, \tau), \\ 
& S_{nm,{\lambda}}^{\bm \alpha}({\bm \eta};h){\bm \alpha}_1
=  S_{nm}^{\bm \alpha}({\bm \eta};h){\bm \alpha}_1 + \lambda J({\bm \alpha}, {\bm \alpha}_1), 
\end{align*}
where $J(\bm \alpha, \bm \alpha_1) = \left\langle {\bm \alpha},  {\bm \alpha}_1 \right\rangle_{K}$. 
Denote $\mathcal{L}({\bm \eta}; h)=\mathrm{E}\mathcal{L}_{nm}({\bm \eta}; h)$, $\mathcal{L}_{{\lambda}}({\bm \eta}; h)=\mathrm{E}\mathcal{L}_{nm, {\lambda}}({\bm \eta}; h)$. The first Fr\'{e}chet derivative of $\mathcal{L}({\bm \eta}; h)$ and $\mathcal{L}_{{\lambda}}({\bm \eta}; h)$ with respect to ${\bm \alpha}$ are 
$S^{\bm \alpha}({\bm \eta};h) = \mathrm{E}S_{nm}^{\bm \alpha}({\bm \eta};h)$, $S_{\lambda}^{\bm \alpha}({\bm \eta};h) = \mathrm{E}S_{nm,\lambda}^{\bm \alpha}({\bm \eta};h),$
respectively. Furthermore, the second Fr\'{e}chet derivative of $\mathcal{L}_{{\lambda}}({\bm \eta}; h)$ with respect to ${\bm \alpha}$ at ${\bm \eta}_{0}$ is
$$
\mathcal{D}S^{\bm \alpha}_{\lambda}({\bm \eta}_{0};h){\bm \alpha}_1 {\bm \alpha}_2  = \mathcal{D}S^{\bm \alpha}({\bm \eta}_{0};h){\bm \alpha}_1 {\bm \alpha}_2 + \lambda J({\bm \alpha}_1, {\bm \alpha}_2),
$$
where 
$
\mathcal{D}S^{\bm \alpha}({\bm \eta}_{0};h){\bm \alpha}_1 {\bm \alpha}_2  = \int_{\mathcal{S}} {\bm \alpha}_1(s, \tau)^T \mathrm{E}\left[f_{e}( 0, s| {\bv}){\bW}_{\bgamma_0}{\bW}_{\bgamma_{0}}^{T}\right] {\bm \alpha}_2(s, \tau) \pi(s) ds, 
$
and $\pi(s)$ is the density function of $\{s_j\}_{j=1}^m$.
To derive the asymptotic theories of the proposed estimators, we define the inner product for any ${\bm \alpha}_1, {\bm \alpha}_2 \in \mathcal{H}^{p+d}$ as
\begin{align*}
&\langle {\bm \alpha}_1, {\bm \alpha}_2 \rangle_{\lambda} = \mathcal{D}S^{\bm \alpha}_{\lambda}({\bm \eta}_{0};h){\bm \alpha}_1 {\bm \alpha}_2,
\end{align*}
and the corresponding norm is denoted as $\|\cdot\|_{\lambda}$.
Moreover, we define a bilinear operator $V(\cdot,\cdot)$ in $\mathcal{H}^{p+d}$ as $V(\bm \alpha_1, \bm \alpha_2)= \int_{\mathcal{S}} {\bm \alpha}_1(s, \tau)^T \mathrm{E}\left[f_{e}( 0, s| {\bv}){\bW}_{\bgamma_0}{\bW}_{\bgamma_0}^{T}\right] {\bm \alpha}_2(s, \tau) \pi(s) ds,$
which implies that $$\langle {\bm \alpha}_1, {\bm \alpha}_2 \rangle_{\lambda}  = V(\bm \alpha_1, \bm \alpha_2) + J(\bm \alpha_1, \bm \alpha_2).$$ 
Similar definition can refer to \cite{yuan2011} and  \cite{shang2015}.

Let $\mathcal{R}_{\lambda}(s_1, s_2)$ be the reproducing kernel matrix of $\mathcal{H}^{p+d}$ endowed with norm $\|\cdot\|_{\lambda}$. Hence, for any ${\bm \theta} \in \mathcal{H}^{p+d}$, ${\bm c} \in \mathbb{R}^{p+d}$ and $s_1, s_2  \in \mathcal{S}$, $(\mathcal{R}_{\lambda, s_1}{\bm c})(s_2) = \mathcal{R}_{\lambda}(s_1,s_2){\bm c}$ and $\langle \mathcal{R}_{\lambda, s_1}{\bm c}, {\bm \theta} \rangle_{\lambda} = {\bm c}^T{\bm \theta}(s_1)$. Additional properties of the vector-valued RKHS are discussed in \cite{minh2016}. Let $P_{\lambda}$ be a positive definite self-adjoint operator $P_{\lambda}: \mathcal{H}^{p+d} \rightarrow \mathcal{H}^{p+d}$ such that $\langle P_{\lambda}\bm \alpha_1, \bm \alpha_2 \rangle_{\lambda} = \lambda J(\bm \alpha_1, \bm \alpha_2)$.	
The following assumption are required to establish the consistency of the proposed estimator $\hat{\bm \eta}$.

\begin{assumption}\rm{
\begin{enumerate}[(i)]
	\item For almost every ${\bz}_{2i}$, the density of $z_{1i}$ conditional on ${\bz}_{2i}$ is everywhere positive; $\sup_i\|{\bv}_i\|_2 < \infty$ almost surely; the true parameter $\bgamma_0$ is in the interior of compact subspaces $\Theta$.
	
	\item The sequence $\{e_i(s, \tau) : s \in \mathcal{S} \}$ is a stochastic process whose $\tau$th quantile conditional on $({\bm v}_i,s,\tau)$ equals zero. Moreover, $\{e_i(s, \tau)\}$ is independent with $z_1$ and $|f_{e}^{(1)}(u,s|\bv_2)| < \infty$ over $(u, s, \bv_2)$. Without loss of generality, we assume that $\mathcal{S}=[0, 1]$
	
	\item The grid points $\{s_j, j = 1, . . . , m\}$ are randomly generated from a density function $\pi(s)$. Moreover, $\pi(s)>0$ for all $s \in [0,1]$ and $\pi(s)$ has continuous second-order derivative with the bounded support $[0,1]$.
 
	\item The minimum eigenvalue of $\mathrm{E}\{f_e(0,s | \bv){\bm W}^{*\otimes 2} | {\bz}\}$ is bounded away from zero uniformly over ${\bz}$.
	
	\item There exists a sequence of functions $\{{\bm \varphi}_l\}_{l\geq 1} \subset \mathcal{H}^{p+d}$ such that $\sup_l\sup_s|{\bm \varphi}_l(s) | < \infty$, and $V({\bm \varphi}_\nu, {\bm \varphi}_l) = \delta_{\nu l}$, $J({\bm \varphi}_\nu, {\bm \varphi}_l)=\rho_l^{-1}\delta_{\nu l}$ for any $\nu, l \geq 1$, where $\delta_{\nu l}$ is the Kronecker delta, and $\rho_{l} \asymp l^{-2r}$. Furthermore, any $\bm \alpha \in \mathcal{H}^{p+d}$ admits the expansion $\bm \alpha = \sum_{l=1}^{\infty}V(\bm \alpha, {\bm \varphi}_l){\bm \varphi}_l$ with convergence in $\mathcal{H}^{p+d}$ under the norm $\|\cdot\|_{\lambda}$.
\end{enumerate}}
\end{assumption}

Assumption 1(\romannumeral1) is employed to establish the asymptotic equivalence between the smoothed and non-smoothed objective functions as $h \rightarrow 0$ \citep{seolinton2005,zhangyy2021}. The boundedness of $\bv_i$ is assumed for theoretical simplicity. Assumption 1(\romannumeral2) assumes that $e_i(s,\tau)$ is independent with $z_1$ so that $f_{e|\bv_2}(u,s |\bv_2) =f_{e|\bv}(u,s |\bv)$ to facilitate the proof, similar assumption can refer to \cite{zhangyy2021}. Assumptions 1(\romannumeral3)-(\romannumeral4) are standard regularity conditions commonly found in the functional data analysis literature \citep{zhu2012,zhou2023}. Assumption 1(\romannumeral5) provides a sequence of basis functions in $ \mathcal{H}^{p+d}$ that simultaneously diagonalizes the operators $V(\cdot, \cdot)$ and $J(\cdot, \cdot)$. Such a diagonalization assumption is common in the literature, as noted in \cite{cai2011aos} and  \cite{shang2015}. 

Let $N = \lambda^{1/(2r)}$, where $r$ is specified in Assumption 1(\romannumeral5). Define a norm $\|\bm \eta(\tau)\|_1 = \|\bm \alpha(\tau)\|_{\lambda} + \|\bgamma(\tau)/h\|_2$. 
The following asymptotic results hold as $n,m \rightarrow \infty$.

\begin{theorem}\label{th1}
	Suppose that Assumption 1(\romannumeral1)-(\romannumeral5) hold, if $n^{-1/2}\lambda^{-1/4-1/(2r)+1/(16r^2)} = o(1)$, then $\|\hat{\bm \eta}(\tau) - {\bm \eta}_{0}(\tau)\|_{1} = O_p((nmN)^{-1/2} + n^{-1/2} + \lambda^{1/2})$.
\end{theorem}

With consistency established, the next step is to derive the limiting distribution. 
Following the approach of \cite{seolinton2005} and \cite{zhangyy2021}, we apply certain linear transformations to the covariates for convenience.  
Let $q_i = z_{1i} + {\bz}_{2i}^T {\bm \gamma}_{0}(\tau)$, and ${\bv}_{2i}$ denote the variables in ${\bv}_i$ excluding $z_{1i}$, then there is a one-to-one relation between $(q_i, {\bv}_{2i})$ and ${\bv}_{i}$, which implies that there exists $(\dot{\theta}_{10}(s, \tau), \dot{\bm \theta}_{20}(s, \tau)^T)^T$ such that $\tilde{\bx}_i^T {\bm \theta}_0(s, \tau) = q_i\dot{\theta}_{10}(s, \tau)+{\bv}_{2i}^T\dot{\bm \theta}_{20}(s, \tau)$. Denote $f_{q|{\bv}_2}(q|{\bv}_2)$ as the density of $q$ conditional on ${\bv}_2$ and $f^{(i)}_{x|{\bv}_2}(x|{\bv}_2) = \partial^if(x|{\bv}_2)/\partial x^i$ as the $i$th order derivative function. 

\begin{assumption}\rm{
\begin{enumerate}[(i)]	
	\item The conditional distribution function $F_e(a, b, s, t | \bv)$ has bounded continuous first and second order partial derivatives and mixed derivatives with respect to $a$ and $b$ for all $s, t \in [0, 1]$ at the given $\tau \in (0, 1)$.
	
	\item $f^{(k)}_{q|{\bv}_2}(q|{\bv}_2)$ is a continuous function and $|f^{(k)}_{q|{\bv}_2}(q|{\bv}_2)| < \infty$ over $(q, {\bv}_2)$ for each integer $0 \leq k \leq k'$($k'$ will be defined later).
	
	\item The smooth function $G$ is twice differentiable with $G(s) + G(-s)=1$; $|G'(w)|$ and $|G''(w)|$ are uniformly bounded over $w$; $\int |G'(w)| dw < \infty$ and $\int |G''(w)| dw < \infty$. Moreover, $\int w \{G(w)-I(w \geq 0)\} dw < \infty$.
 
	\item For each integer $1\leq j \leq k'$, $\int w^{j-1}\{G(w)-I(w \geq 0)\}G'(w)=0$ and $\int w^{k'}\{G(w)-I(w\geq0)\}G'(w) \neq 0$.
 
	\item $nh^3 \rightarrow 0$ if $k'=1$ and $nh^4 \rightarrow 0$ if $k' > 1$. In addition, $nh^2 \rightarrow \infty$.
	\item The eigenvalues of $\mathbb{V}(\tau)$ (defined later) are bounded below and above by some positive constants $c_1$ and $1/c_1$, respectively.
	The eigenvalues $\mathbb{Q}(\tau)$ (defined later) are bounded below and above by some positive constants $c_2$ and $1/c_2$, respectively.
\end{enumerate}}
\end{assumption}

Assumption 2(\romannumeral1)  is a standard condition in the quantile functional data analysis literature \cite{zhou2023}. Assumption 2(\romannumeral2) imposes constraints on the conditional density of $q_i$. Assumption 2(\romannumeral3)-(\romannumeral4) pertain to the smoothing function $G(\cdot)$. Assumption 2(\romannumeral5) relates to the bandwidth parameter $h$. Assumption 2(\romannumeral2)-(\romannumeral6) are commonly found in the change-plane literature \cite{seolinton2005} and  \cite{zhangyy2021}.

\begin{theorem}\label{th2}
	Suppose that Assumptions in Theorem \ref{th1} and Assumption 2 (\romannumeral1)-(\romannumeral6) hold, if 
	$N=o(1)$, $hN^{-1} = o(1)$, $m^{-1}N^{-2}h= o(1)$, $nhN^{2r-1}=o(1)$, $n^{-1}N^{-1}=o(1)$, and $\sum_{l=1}^{\infty}\rho_l^2V(\bm \alpha_{0}, \bm \varphi_l) < \infty$, we have $\sqrt{n}\Lambda(s,\tau)^{-1/2}\left(\hat{\bm \alpha}(s,\tau)-{\bm \alpha}_{0}(s,\tau)\right) \stackrel{d}{\rightarrow}  N(0, {\bf I}_{p+d})$, $\sqrt{n/h}\left(\hat{\bgamma}(\tau)-{\bgamma}_{0}(\tau)\right) \stackrel{d}{\rightarrow}  N(0,\\ \mathbb{Q}(\tau)^{-1}\mathbb{V}(\tau)\mathbb{Q}(\tau)^{-1})$,
	and they are asymptotically independent, where $\stackrel{d}{\rightarrow}$ represents convergence in distribution,
	\begin{align*}
	&{\Lambda}(s,\tau) = \int_0^1 \int_0^1 \mathrm{E}_{\bm v} \left\{[F_e(0,0,t_1, t_2 |{\bm v}) - \tau^2]\bW_i^{*T}\mathcal{R}_{\lambda}(s, t_1)\mathcal{R}_{\lambda}(s, t_2) \bW_i^*\right\}\pi(t_1)\pi(t_2)dt_1dt_2,\\
	&\mathbb{V}(\tau) = \int G'(\xi)^2 d\xi \int_0^1 \int_0^1\mathrm{E}_{\bv_2}\left[(F_{e}(0, 0, s, t \mid {\bv}) - \tau^2) \dot{\bm \theta}_{20}(s)^T{\bv_2}{\bv_2}^T\dot{\bm \theta}_{20}(t){\bz}_{2}{\bz}_{2}^T f_{q|{\bv_2}}(0 |{\bv_2}) \right]\pi(s)\pi(t)dsdt,\\
	&\mathbb{Q}(\tau) = G'(0)\int_0^1 \mathrm{E}_{\bv_2}\left\{f_e(0,s|\bv_2)(\bv_2^T\dot{\bm \theta}_{20}(s))^2\bz_{2}\bz_{2}^Tf_{q|\bv_2}(0|\bv_2)\right\}\pi(s)ds.
	\end{align*}
\end{theorem}

\begin{remark}{\rm
As noted in the change-plane regression literature, there is asymptotic independence between the functional coefficients ${\bm \alpha}(s,\tau)$ and the grouping parameter ${\bgamma(\tau)}$.
The convergence rate of ${\bgamma(\tau)}$ is $\sqrt{h/n}$, which is consistent with the results found in change-plane analysis of scalar data \citep{seolinton2005,su2019,zhangyy2021}. Consequently, statistical inference for the functional coefficients ${\bm \alpha}(s,\tau)$ can be conducted as if ${\bgamma(\tau)}$ were known.
}
\end{remark}

\begin{remark}{\rm 
	Theorem \ref{th2} enables us to construct a pointwise $(1-\zeta) \times 100\%$ confidence bands of ${\bm \alpha}_0(s,\tau)$:
	$[\hat{\bm \alpha}(s,\tau) - q_{1-\zeta/2}\Lambda(s,\tau), \hat{\bm \alpha}(s,\tau) + q_{1-\zeta/2}\Lambda(s,\tau)],$
	where $q_{1-\zeta/2}$ is the $(1-\zeta/2)$-th quantile of a standard normal distribution.}
\end{remark}

\section{Subgroup testing}\label{sec3}
In this subsection, we introduce a novel test statistic for subgroup testing in functional data. In essence, the subgroup testing is equivalent to the following hypothesis test problem:
\begin{gather*}
H_{0}: {\bm \theta}(s,\tau)  \equiv 0, \forall s \in[0,1]\quad \text{ versus } \quad  H_{1}: {\bm \theta}(s,\tau) \neq 0 , \exists s \in[0,1].
\end{gather*}
It is evident that inference for change-plane models is challenging, as the grouping parameter $\bgamma(\tau)$ is not identifiable under the null hypothesis.
Some widely used methods \citep[e.g.,][]{songsui2017,huang2020} establish the test statistic by searching for the supremum value over the parametric space of $\bgamma(\tau)$, which becomes challenging when the dimension of the parametric space is high. Inspired by \cite{liu2024}, we propose a weighted average of squared score statistics to delineate the parameter $\bgamma(\tau)$, thereby reducing computational burdens.

\subsection{Weighted average of squared score statistic}
Denote $\{{\bm u}_i(s)=\left(y_i(s), \bm x_i, \tilde{\bm x}_i, \bm z_i\right), s \in [0,1]\}_{i=1}^n$
as the $n$ copies of ${\bm u}(s) = \left(y(s), \bm x, \tilde{\bm x}, \bm z \right)$.
For a fixed ${\bm \gamma} \in \mathbb{R}^q$ and a given $s$, we obtain an estimating equation of $\bm \theta(s,\tau)$ under the null hypothesis as follows:
\begin{align}
\sum_{i=1}^{n}\{I(y_{i}(s) - {\bm x}_{i}^{T} \hat{\bm \beta}(s,\tau) \leq 0)-\tau\}\tilde{\bm x}_{i}I(z_{1i} + {\bz}_{2i}^T{\bgamma} \geq 0) = 0,
\label{eq:score}
\end{align}
where $\hat{\bm \beta}(s,\tau)$ is an estimator of ${\bm \beta}(s,\tau)$ under the null hypothesis according to:
\begin{align}\label{eq3.2}
\hbbeta =\underset{\bbeta \in \mathcal{H}^p}{\arg\min}\frac{1}{nm}\sum_{i=1}^{n}\sum_{j=1}^{m}
\rho_{\tau}\left (y_i(s_j)- \bx_i^{T}\bbeta(s_j,\tau)\right)
+\frac{\lambda}{2}J(\bbeta,\bbeta),
\end{align}
where $\mathcal{H}^p$ is the $p$-dimensional vector RKHS.
Denote $\psi_1({\bm u}_i(s),{\bm \beta}(s,\tau),0,{\bm \gamma})=[I(y_{i}(s) - {\bm x}_{i}^{T} {\bm \beta}(s,\tau) \leq 0)-\tau]\tilde{\bm x}_{i}I(z_{1i} + {\bz}_{2i}^T{\bgamma} \geq 0)$, and $\psi_2({\bm u}_i(s),\bm \beta(s,\tau)) = m^{-1}\sum_{j=1}^m \{I(y_i(s_j) \leq {\bm x}_i^T {\bm \beta}(s_j,\tau))-\tau\}K(s_j,s){\bm x}_i$.

Based on the estimation equation (\ref{eq:score}), we consider the following weighted average squared score test statistic (WAST):
\begin{align}\label{eq:Tn}
T_n = \frac{1}{mn(n-1)}\sum_{k=1}^m\sum_{i \ne j}w_{ij}\tbx_i^{T}\tbx_j\left[ I(y_i(s_k) \leq \bx_i^{T}\hat{\bbeta}(s_k,\tau)) - \tau \right]
\left[ I(y_j(s_k) \leq \bx_j^{T}\hat{\bbeta}(s_k,\tau)) - \tau \right],
\end{align}
where 
\begin{align}
w_{ij} = \frac{1}{4}+\frac{1}{2\pi}\arctan\left(\frac{\varrho_{ij}}{\sqrt{1-\varrho_{ij}^2}}\right) \quad if \quad i\ne j,
\label{eq:w}
\end{align}
is a weight for removing the nuisance parameter $\bgamma$, and $\varrho_{ij}=\bz_i^{T}\bz_j\left(\left\lVert  \bz_i\right\rVert\left\lVert \bz_j \right\rVert  \right)^{-1}$. 
It is clear that the proposed statistic $T_n$ has a closed expression which reduces the burden of computation.

\begin{remark}{\rm 
	The proposed test statistic (\ref{eq:Tn}) can be regarded as a natural extension of the weighted average squared score test statistic of \cite{liu2024} to functional data, that is, one can equivalently write $T_n$ in the following form:
	\begin{align*}
	T_n  = \frac{1}{mn(n-1)}\sum_{k=1}^m\sum_{i \ne j}\rho_{ij}(s_k,\tau),
	\end{align*}
	where
	$\rho_{ij}(s_k,\tau) =\int_{\bm \gamma \in \Theta} w(\bm \gamma)\psi_1({\bm u}_i(s_k),\hat{\bm \beta}(s_k,\tau),0,\bm \gamma)^T \psi_1({\bm u}_j(s_k),\hat{\bm \beta}(s_k,\tau),0,\bm \gamma)d\bm \gamma,$
	and $w(\bm \gamma)$ is a weight satisfying $w(\bm \gamma) \geq 0$ for all $\bm \gamma \in \Theta$, $\int_{\bm \gamma \in \Theta}w(\bm \gamma)d\bm \gamma=1$. 
	As noted in \cite{liu2024}, the density $w(\bm \gamma)$ can be considered as a prior distribution for the grouping parameter $\bm \gamma$. Although the choice of weight does not affect the asymptotic distributions, it can influence the computation of the test statistic due to the numerical integration over the nuisance parameter $\bm \gamma$.
	By selecting the distribution in (\ref{eq:w}), the statistic $T_n$ can be expressed in closed form as shown in (\ref{eq:Tn}).
}
\end{remark}

\subsection{Limiting distributions of test statistic}\label{sec3.2}
In this subsection, we establish the asymptotic properties of the proposed statistic under both the null hypothesis and the local alternative hypothesis. For simplicity, we assume that the sampling points $\{s_j\}_{j=1}^m$ are uniformly distributed over $[0, 1]$.

For convenience, we introduce some additional notations. For a given $\tau$-th quantile, define the kernel of a U-statistic under the null hypothesis as
\begin{align*}
h({\bm u}_i(s),{\bm u}_j(s))=& \int_{\bm \gamma \in \Theta} \psi_1({\bm u}_i(s),{\bm \beta}_{0}(s,\tau),0,\bm \gamma)^{T}\psi_1({\bm u}_j(s),{\bm \beta}_0(s,\tau),0,\bm \gamma)w(\bm \gamma)d\bm \gamma \\
&+ \psi_2({\bm u}_i(s),{\bm \beta}_0(s,\tau))^{T}K_j(s) + K_i(s)^{T}\psi_2({\bm u}_j(s),{\bm \beta}_0(s,\tau)) \\
&+ \psi_2({\bm u}_i(s),{\bm \beta}_0(s,\tau))^{T}H(s)\psi_2({\bm u}_j(s),{\bm \beta}_0(s,\tau)),
\end{align*}
where 
\begin{align*}
K_i(s)  = & \int_{\bm \gamma \in \Theta}U_2(s)^{T}U_1(s,\bm \gamma)^{T}\psi_1({\bm u}_i(s),{\bm \beta}_0(s,\tau),0,\bm \gamma)w(\bm \gamma)d\bm \gamma, \\
H(s) = & \int_{\bm \gamma \in \Theta} U_2(s)^{T}U_1(s,\bgamma)^{T}U_1(s,\bm \gamma)U_2(s)w(\bm \gamma)d\bm \gamma,
\end{align*}
with $U_1(s,{\bm \gamma}) = \partial \mathrm{E}\psi_1({\bm u}_i(s),{\bm \beta}_0(s,\tau),0,{\bm \gamma})/\partial {\bm \beta}^T$ and $U_2(s) = \{\partial{\mathrm{E}\psi_2(\bm u_i(s),\bbeta_0(s,\tau))}/\partial{\bbeta^T}\}^{-1}$.
Denote ${\bm \Sigma}_{\psi_1}(s) = \mathrm{E}\{[I(e_i(s, \tau) \leq 0)-\tau]^2 \tbx_i^{\otimes 2}\}$ and ${\bm \Sigma}_{\psi_2}(s) = \mathrm{E}\{\psi_2(\bm u_i(s), \bbeta_0(s,\tau))^{\otimes 2}\}$.
To obtain the asymptotic results, we need following regular assumptions. 

\begin{assumption}\rm{
\begin{enumerate}[(i)]	
	\item $0< \mathrm{E}\{I(z_{1i} + {\bz}_{2i}^T{\bgamma} \geq 0)\} <1$ for any $\bgamma \in \Theta$.
	\item For any $s$, $\left[\mathrm{E}\{f_e(0, s|\bv) \bx_i \tbx_i^{T}\}\right]^{\otimes 2}$, $U_2(s)$, ${\bm \Sigma}_{\psi_1}(s)$, and ${\bm \Sigma}_{\psi_2}(s)$ are finite and positive definite matrix.
	\item There is a positive function $b(\bm u_i(s), {\bm \delta}(s,\tau))$ of $\bm u_i(s)$ which relies on $\bbeta_0(s,\tau), \bgamma_0(\tau)$ such that 
	$$\left|{\bm \delta}(s,\tau)^T 
	\frac{\{\partial f(\bm u_i(s), \bbeta_0(s,\tau),a_n{\bm \delta}(s,\tau),\bgamma_0(\tau))/\partial {\bm \theta} \}}{f(\bm u_i(s),\bbeta_0(s,\tau),0,\bgamma_0(\tau))} \right| \leq b(\bm u_i(s), {\bm \delta}(s,\tau)),$$ 
	and $\mathrm{E}[b(\bm u_i(s),{\bm \delta}(s,\tau))^2]$, $\lambda_{\max}(\mathrm{E}[b(\bm u_i(s), {\bm \delta})\psi_1(\bm u_i(s),\bbeta_0(s,\tau),{\bm \theta}(s,\tau),\bgamma_0(\tau))^{\otimes 2}])$, and \\ $\lambda_{\max}(\mathrm{E}[b(\bm u_i(s), {\bm \delta}(s,\tau))\psi_2(\bm u_i(s),\bbeta_0(s,\tau))^{\otimes 2}])$ are bounded by $C_f({\bm \delta}(s,\tau))$, where $\lambda_{\max}(\cdot)$ is the maximum eigenvalue, and for all $k$, $\mathrm{E}[\phi_k(\bm u_i(s))^2b(\bm u_i(s),{\bm \delta}(s,\tau))]$ is bounded by $C_f({\bm \delta}(s,\tau))$, where $a_n = o(1)$, $C_f({\bm \delta}(s,\tau))>0$ relying on ${\bm \delta}(s,\tau)$, $\phi_k(\cdot)$ is defined in Theorem \ref{th5}, and $\bm u(s)$ is generated from the null distribution with density $f(\bm u(s),\bbeta_0(s,\tau),0,\bgamma_0(\tau))$.
\end{enumerate}}
\end{assumption}
Assumption 3(\romannumeral1) is satisfied for many commonly used change-plane models \cite{songsui2017} and  \cite{su2020}. Assumption 3(\romannumeral2)-(\romannumeral3) are also assumed in the inference of change-plane \cite{liu2024}.

\begin{theorem}\label{th4}
	If Assumptions in Theorem \ref{th2} and Assumption 3(\romannumeral1)-(\romannumeral2)  hold, then under the null hypothesis, we have
	$$nT_n - \mu_0 \stackrel{d}{\rightarrow} \int_{0}^1 \xi(s) ds,$$
	where $\mu_0=\int_{0}^1 \{2\mathrm{E}[\psi_2({\bm u}_i(s),{\bm \beta}_0(s,\tau))^{T}K_i(s)] + \mathrm{E}[\psi_2({\bm u}_i(s),{\bm \beta}_0(s,\tau))^{T}H(s)\psi_2({\bm u}_i(s),{\bm \beta}_0(s,\tau))]\}ds,$
	$\xi(s) = \sum_{k=1}^\infty \lambda_k(s)(\chi_{1k}^2-1)$, and $\chi_{1k}^2$ are independent $\chi_1^2$ random variables, i.e., $\xi(s)$ has the characteristic function $\mathrm{E}[e^{it\xi(s)}]=\prod_{k=1}^\infty(1-2it\lambda_k(s))^{-1/2}e^{-it\lambda_k(s)},$
	where $i=\sqrt{-1}$ is the imaginary unit, and $\{\lambda_k(s)\}$ are the eigenvalues of kernel $h({\bm u}_1(s),{\bm u}_2(s))$ under $f(\bm u(s),\bbeta_0(s,\tau),0,\bm \gamma_0(\tau))$, i.e., they are the solution of $\lambda_k(s)g_k(\bm u_2(s))=\int_0^\infty h({\bm u}_1(s),{\bm u}_2(s))g_k(\bm u_1(s))f({\bm u}_1(s),\bm \beta_0(s,\tau),0,\bm \gamma_0(\tau))d\bm u_1(s)$ for nonzero function $g_k$, where $f(\bm u(s), \bbeta(s,\tau), \bm \theta(s,\tau),\bm \gamma(\tau))$ is the density of $\bm u(s)$ with parameters $\bbeta,\bm \theta$, and $\bgamma$.
\end{theorem}

\begin{remark}{\rm 
	Theorem \ref{th4} demonstrates that $T_n$ is a U-statistic. Under the null hypothesis, if $\bbeta(\cdot,\tau) = \bbeta_0(\cdot,\tau)$ is known, the bias $\mu_0$ approaches zero.}
\end{remark}

Next, we investigate the power performance of the proposed test statistic under the local alternative $H_{1n}: \bm \theta(s,\tau)=n^{-1/2}\bm \delta(s,\tau)$. 

\begin{theorem}\label{th5}
	If Assumptions in Theorem \ref{th2} and Assumption 3 (\romannumeral1)-(\romannumeral3) hold, then under $H_{1n}$, we have
	$$nT_n - \mu_0 \stackrel{d}{\rightarrow} \int_{0}^1 \xi(s)ds,$$
	where $\mu_0$ is defined in Theorem \ref{th4}, $\xi(s)$ is a random variable of the form $\xi(s) = \sum_{k=1}^\infty \lambda_k(s)(\chi_{1k}^2(\mu_{ak}(s))-1)$, and $\chi^2_{1k}(\mu_{ak}(s))$ are independent noncentral $\chi_{1}^2$ variables, i.e., $\xi(s)$ has the characteristic function $\mathrm{E}[e^{it\xi(s)}]=\prod_{k=1}^\infty(1-2it\lambda_k(s))^{-1/2}e^{-it\lambda_k(s) + \{{it\lambda_k(s)\mu_{ak}(s)}/{1-2it\lambda_k(s)}\}},$
	where $\{\lambda_k(s)\}$ are the eigenvalues of kernel $h({\bm u}_1(s),{\bm u}_2(s))$ under $f({\bm u}(s),\bm \beta_0(s,\tau),0,\bm \gamma_0(\tau))$, i.e., they are the solution of $\lambda_k(s)g_k(\bm u_2(s))=\int_0^\infty h({\bm u}_1(s),{\bm u}_2(s))g_k(\bm u_1(s))f({\bm u}_1(s),\bm \beta_0(s,\tau),0,\bm \gamma_0(\tau))d\bm u_1(s)$ for nonzero function $g_k$, and each noncentrality parameter of $\chi^2_{1j}(\mu_{ak}(s))$ is 
	$$\mu_{ak}(s)= \mathrm{E}[\phi_k(\bm u_0(s))\bm \delta(s)\{\partial \log(f({\bm u}_0(s),\bm \beta_0(s,\tau),0,\bm \gamma_0(\tau)))/\partial {\bm \theta}\}], \quad k=1,2,\cdots,$$
	where $\phi_k(\bm u(s))$ denotes orthonormal eigenfunctions corresponding to the eigenvalues $\{\lambda_k(s)\}$ and ${\bm u}_0(s)$ is generated from the null distribution $f({\bm u}(s),\bm \beta_0(s,\tau),0,\bm \gamma_0(\tau))$.
\end{theorem}
\begin{remark}{\rm
According to Theorem \ref{th5}, the power function can be theoretically approximated using the distribution of $\xi(s)$ for $s \in [0,1]$. The additional noncentrality parameters $\{\mu_{ak}(s), s \in [0,1]\}$ in Theorem \ref{th5} quantify the discrepancy between the null and alternative hypotheses, with the power increasing as the alternative hypothesis diverges further from the null.		
}
\end{remark}

\subsection{Calculation of the critical value}
Given the asymptotic results of the proposed test statistic in Subsection \ref{sec3.2}, calculating the $\alpha$-th quantile of the asymptotic distribution directly is challenging. Inspired by \cite{songsui2017} and \cite{liu2024}, we propose a bootstrap method to approximate the critical value or p-value.

Let $\left\{{\bm u}_i^{*}(s_k)=\left(y_i^{*}(s_k), \bm x_i, \tilde{\bm x}_i, \bm z_i\right), k=1,\dots,m\right\}_{i=1}^n$ be the bootstrap samples generated from the distribution $f({\bm u}_i(s), \hat{\bm \beta}(s,\tau), 0 , \bm \gamma)$ where $\hat{\bm \beta}(s,\tau)$ is obtained by (\ref{eq3.2}). We calculate the statistic $T_n^{*}$ based on the bootstrap samples by
$$T_n^{*} = \frac{1}{mn(n-1)}\sum_{k=1}^m\sum_{i \ne j}\int_{\bm \gamma} \psi_1({\bm u}_i^{*}(s_k),\hat{\bm \beta}^{*}(s_k,\tau),0,\bm \gamma)^T \psi_1({\bm u}_j^{*}(s_k),\hat{\bm \beta}^{*}(s_k,\tau),0,\bm \gamma)w(\bm \gamma)d\bm \gamma,$$
where $\hat{\bm \beta}^{*}(s_k,\tau)$ is obtained from (\ref{eq3.2}) by using the bootstrap sample under the null hypothesis. 

Repeat the generation procedure $B$ times, where $B$ is a sufficiently large integer. We obtain bootstrap samples $\{{\bm u}_i^{*b}(s_k)=\left(y_i^{*b}(s_k), \bm x_i, \tilde{\bm x}_i, \bm z_i\right), k=1,\dots,m\}_{i=1}^n$ and compute $T_n^{*b}$ based on each $b$-th iteration, with $b = 1, \cdots, B$.
The critical value $C_{\alpha}$ can be determined from the empirical distribution of $\{T_n^{*b}\}_{b=1}^B$, where $C_{\alpha}$ is the upper $\alpha$-th quantile of this empirical distribution. The p-value value is approximated by $B^{-1}\sum_{b=1}^B I(T_n > T_n^{*b})$. Specifically, the bootstrap procedure is presented in Algorithm \ref{alg2}.

\begin{algorithm}[hbt!]\footnotesize
	\caption{Calculate the p-value of $T_n$ by the bootstrap.}
	\label{alg2}
	\begin{algorithmic}[1]
		\STATE \textbf{Input:} Datasets $\{(y_i(s_k),\bx_i,\tbx_i,\bz_i), k =1,\cdots m\}_{i=1}^n$, quantile level $\tau$, bootstrap iterations $B$.
		\STATE Step {\bf 1}: Calculate the statistic $T_n$ by (\ref{eq:Tn}) based on the observed datasets.
		\STATE Step {\bf 2}: Obtain the estimated errors by $\hat{e}_i(s_k,\tau) = y_i(s_k) - \hat{y}_i(s_k)$ for $i=1,\cdots, n$, where $\hat{y}_i(s_k) = \bx_i^{T}\hat{\bbeta}(s_k,\tau)$ and $\hbbeta(s_k,\tau)$ is obtained by (\ref{eq3.2}).
		\STATE Step {\bf 3}: Calculate the p-value by the bootstrap method.
		\FOR{$b = 1, 2, \cdots, B$}
		\STATE Step 3.1: Construct bootstrap samples by $y_i^{*b}(s_k)=\hat{y}_i(s_k)+w_{i}^{*b}| \hat{e}_i(s_k,\tau)|$ for $i=1,\cdots, n$, where the weight $w_i^{*b}$ following a discrete distribution with $P(w_i^{*b}=2(1-\tau))=1-\tau$ and $P(w_i^{*b}=-2\tau)=\tau$.
		\STATE Step 3.2: Calculate the statistic ${T^{*b}_n}$ based on bootstrap samples $ \{(y_i^{*b}(s_k),\bx_i,\tbx_i,\bz_i ), k =1,\cdots m\}_{i=1}^n$.
		\ENDFOR
		\STATE \textbf{Output:}  The p-value $p=\sum_{b=1}^{B}I(T_n > T_n^{*b})/B$.
	\end{algorithmic}
\end{algorithm}

Theorem \ref{th6} below provides the consistency of the bootstrap distribution of the test statistic. The following assumption  is required to derive the asymptotic results.

\begin{assumption}\rm{
There is a positive function $b_1(\bm u_i(s), {\bm \delta}_1(s,\tau))$ of $\bm u_i(s)$ which relies on $\bbeta_0(s,\tau), \bgamma_0(\tau)$ such that 
$$\left|{\bm \delta}_1(s)^T \frac{\{\partial f(\bm u_i(s),\bbeta_0(s,\tau)+a_n{\bm \delta}_1(s,\tau), 0, \bgamma_0(\tau))/\partial {\bm \beta} \}}{f(\bm u_i(s),\bbeta_0(s,\tau),0,\bgamma_0(\tau))}\right| \leq b_1(\bm u_i(s), {\bm \delta}_1(s,\tau)),$$ 
and $\mathrm{E}[b_1(\bm u_i(s),{\bm \delta}_1(s,\tau))^2]$, $\lambda_{\max}(\mathrm{E}[b_1(\bm u_i(s), {\bm \delta})\psi_1(\bm u_i(s),\bbeta_0(s,\tau),{\bm \theta}(s,\tau),\bgamma_0(\tau))^{\otimes 2}])$, and\\ $\lambda_{\max}(\mathrm{E}[b(\bm u_i(s), {\bm \delta}_1(s,\tau))\psi_2(\bm u_i(s),\bbeta_0(s,\tau))^{\otimes 2}])$ are bounded by $C_f({\bm \delta}_1(s,\tau))$, and for all $k,k'$, $\mathrm{E}[\phi_k(\bm u_i(s))^2b_1(\bm u_i(s),{\bm \delta}_1(s,\tau))]$ and $\mathrm{E}[\phi_k(\bm u_i(s))\phi_{k'}(\bm u_i(s))|b_1(\bm u_i(s),{\bm \delta}_1(s,\tau))]$ are bounded by $C_f({\bm \delta}_1(s,\tau))$, where $a_n = o(1)$, $C_f({\bm \delta}(s,\tau))>0$ relying on ${\bm \delta}(s,\tau)$, $\phi_k(\cdot)$ is defined in Theorem \ref{th5}, and $\bm u(s)$ is generated from the null distribution with density $f(\bm u(s),\bbeta_0(s,\tau),0,\bgamma_0(\tau))$.}
\end{assumption}

\begin{theorem}\label{th6}
	If Assumptions in Theorem \ref{th2}, Assumption 3 (\romannumeral1)-(\romannumeral2) and Assumption 4 hold, then under the null hypothesis, we have
	$\sup_{x \in \mathbb{R}}|P(nT_n \leq x) - P^*(nT^*_n \leq x)|  \stackrel{p}{\rightarrow} 0,$
	where $P^*$ denotes the probability under the bootstrap procedure, and $\stackrel{p}{\rightarrow}$ represents covergence in probability.
\end{theorem}

\begin{remark}{\rm
	Theorem \ref{th6} implies that $nT^*_n$ and $nT_n$ have the same asymptotic distribution, providing the asymptotic consistency of the bootstrap distribution for the test statistic.}
\end{remark}

\section{Simulation studies}\label{sec4}
\subsection{Performance of estimation}
\label{sec:estimation}
In this subsection, we demonstrate the finite-sample performance of the proposed estimators under various scenarios. Consider the following functional change-plane quantile regression model:
\begin{align}\label{model:qr4.1}
 Q_{y_{i}(s_j)}(\tau | {\bx}_{i}, \widetilde{\bx}_i, {\bz}_{i}, s_j)={\bx}_{i}^{T} {\bm \beta}(s_j, \tau)+\widetilde{\bx}_i^{T} {\bm \theta}(s_j, \tau) I\left({\bz}_{i}^{T} {\bm \psi}({\tau}) \geq 0\right), 
\end{align} 
with $i = 1, \cdots ,n$ and $j = 1, \cdots, m$. The variables  $\bx_i^{T}=\left(1, \tilde{\bx}_i^{T}\right)$ with $\tilde{\bx}_i=\left(x_{1i},x_{2i}\right)^{T}$ generating from multinormal distribution with mean zero and covariance $\Sigma_{\tilde{\bx}}=(0.5^{|s-k|})$ for $s,k=1,2$, the grouping variables $\bz_{i}=\left(z_{1i}, 1, z_{2i}\right)^{T} $ with  $z_{1i}\sim N(0,1)$, $z_{2i}\sim N(1,1)$, and $\{s_j\} \sim U[0,1]$.
The functions $\bbeta(s,\tau)=(\beta_1(s,\tau),\beta_2(s,\tau),\beta_3(s,\tau))^{T}$and $\bm \theta(s,\tau)=(\theta_1(s,\tau),\theta_2(s,\tau))^{T}$,
where
$\beta_1(s,\tau)=\sin(\pi s)$, $\beta_2(s,\tau) = \left(1-s\right)^3$, $\beta_3(s,\tau)=\exp\left(-3s\right)$, $
\theta_1(s,\tau)=4\cos(0.5\pi s)+3s^3$,  and $\theta_2(s,\tau)=3s^2 + 3$. 
Let ${\bm \psi}(\tau) = (1, {\bm \gamma}^T(\tau))^{T}$ with ${\bm \gamma}(\tau)=\left(-1,1\right)^T.$
Here we consider three scenarios, assuming that  $\left\{\tilde{e}_i(s_j)\right\}$ follows a multivariate normal, $t(3)$, or Laplace distribution.
Moreover, let the multivariate  normal distribution and Laplace distribution share the same zero mean and covariance structure $\Sigma_{\tilde{e}} = (\exp\{-(s_j-s_l)^2/0.8^2\})$ for $j, l =1,\cdots, m$, and let the covariance structure of  $t(3)$ be $3\Sigma_{\tilde{e}}$.
Then let $e_i(s_j, \tau) = \tilde{e}_i(s_j) - F^{-1}(\tau)$, where
$F$ is the marginal density function of $\tilde{e}_i(s_j)$, resulting in the $\tau$-th quantile of $e_i(s_j, \tau)$ equals to zero.
Chose the Gaussian kernel $K(s,t)=\exp \{-{||s-t||_2^2}/({2\sigma^2})\}$ with $\sigma = 0.2$ as the reproducing kernel of $\mathcal{H}$.

Herein, we set the tuning parameters uniformly from the interval $[2,8]$, that is, $\{\tilde{\lambda}_t \in [2,8], t=1,\cdots,40\}$, and let $\lambda_t = \tilde{\lambda}_t/(nm)$ for each $t$. The optimal tuning parameter $\lambda$ is determined by minimizing the mean integrated squared error (RMISE) of estimated functions, which is defined as:
${\rm RMISE}(\lambda) =  \sum_{k=1}^{p+d}[m^{-1}\sum_{j=1}^{m}(\hat{\alpha}_k(s_j)-\alpha_k(s_j))^2 ]^{1/2}$.
To show the accuracy of identifying subgroups, we calculate
$
{\rm Accuracy} = 1-{n}^{-1} \sum_{i=1}^{n}\vert I\left({\bz}_{i}^{T} {\bm \psi}({\tau}) \geq 0\right)-I\left({\bz}_{i}^{T} {\bm \psi}({\tau}) \geq 0\right)\vert .
$
To explore the applicability of the method under different scenarios, we set the quantile levels $\tau=0.25, 0.5, 0.75$, the sample sizes $n =  200, 300, 400$, and $m = 30, 50$. 
The simulations are repeated 500 times under each scenario. 
Due to space constraints, we present only the results for the multivariate $t(3)$ distribution, with the remaining results provided in the Supplementary Material.

\begin{figure}[!ht]
	\centering
	\includegraphics[height=4cm,width=4.5cm]{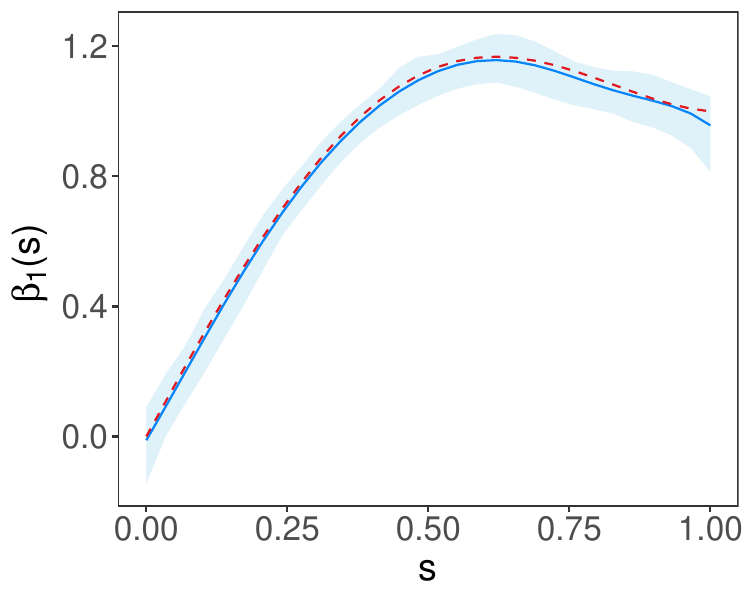}
	\includegraphics[height=4cm,width=4.5cm]{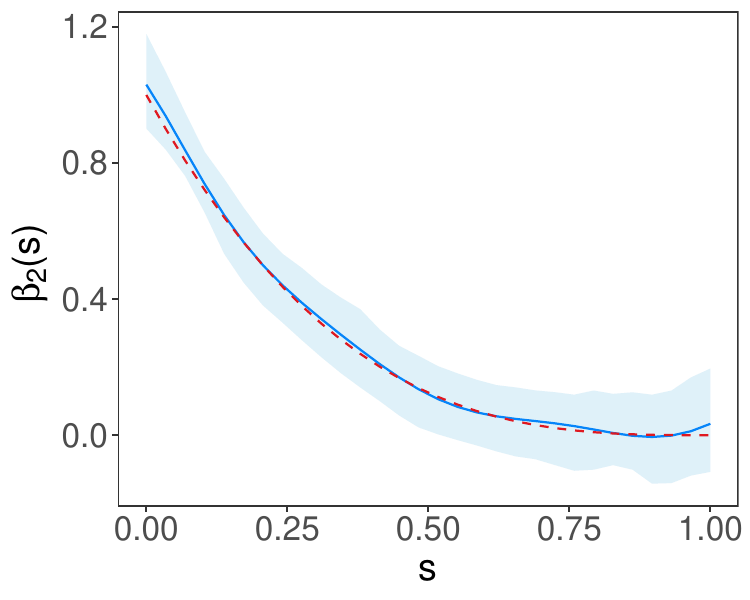}
	\includegraphics[height=4cm,width=4.5cm]{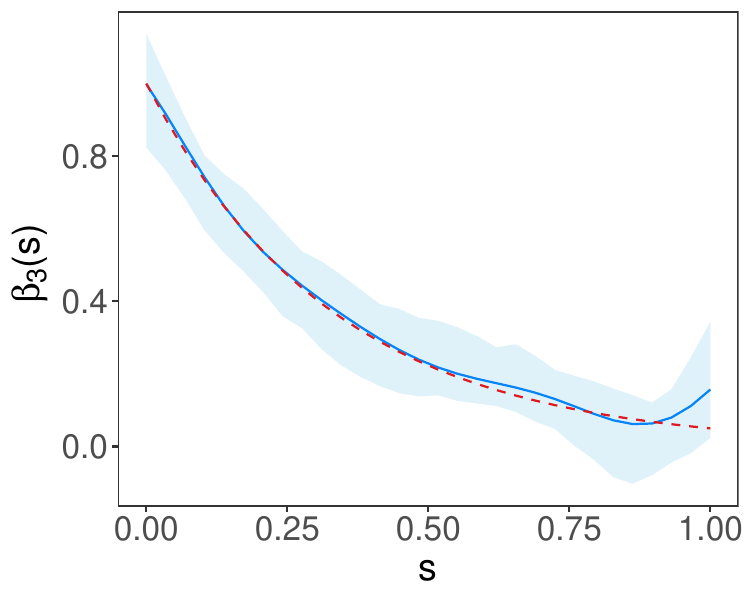}
	\includegraphics[height=4cm,width=4.5cm]{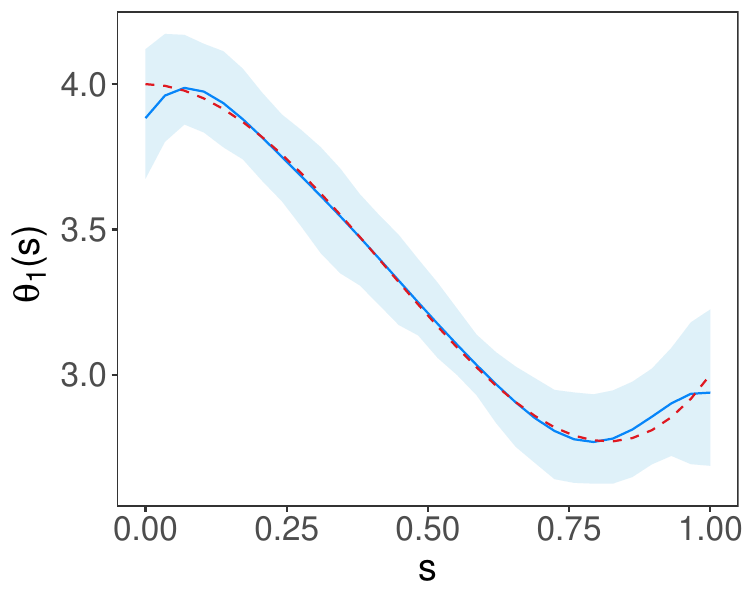}
	\includegraphics[height=4cm,width=4.5cm]{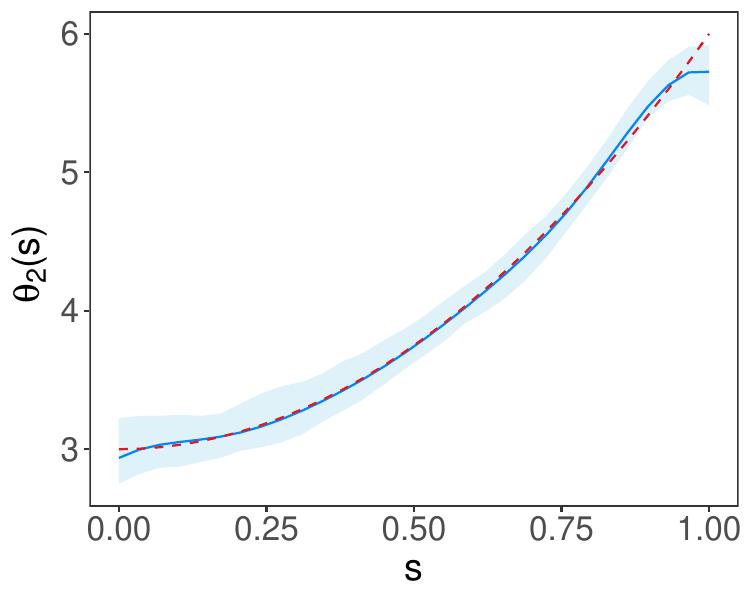}
	\caption{\footnotesize{The true function (dashed line), the mean estimated function (solid line) from 500 repetitions, and the 95\% pointwise confidence bands for each component function at $(\tau,n,m)=(0.75,400,30)$.}}
	\label{fig2:n400m30tau75}
\end{figure}

Table \ref{table2:rmise} lists the RMISE of each component function.  The results indicate that the RMISE of each component function decrease as the sample size $n$ and the observed data points $m$ increase, which verifies the theoretical results shown in Theorems \ref{th1} and \ref{th2}. 
Additionally, the standard deviations are small and exhibit minimal variation, indicating that the proposed estimation method is stable.
Figure \ref{fig2:n400m30tau75} depicts the true function and the mean of estimated functions from 500 repetitions for each component function at $(\tau,n,m)=(0.75,400,30)$.  It is clear that for each component function, the estimated function is close to the true function with narrow confidence bands, which is consistent to the asymptotic results in Theorem \ref{th2}.

\begin{table}[!ht]\tiny
	\renewcommand{\arraystretch}{1.2}
	\caption{The RMISE of each component function and the standard deviation in subscript brackets with $\tau=0.25 ,0.5, 0.75$ and error distribution $t(3)$.}
	\resizebox{\textwidth}{!}{
		\begin{tabular}{cccccccc}
			\hline 
			\multirow{2.5}*{$\tau$}&&\multicolumn{2}{c}{$n=200$}& \multicolumn{2}{c}{$n=300$}& \multicolumn{2}{c}{$n=400$} \\
			\cmidrule(lr){3-4} \cmidrule(lr){5-6} \cmidrule(lr){7-8}
			&& $m=30$ & $m=50$ & $m=30$ & $m=50$ &$ m=30 $& $m=50$   \\ \cmidrule(lr){1-8}
			{$0.25$}& $\beta_1(\cdot)$&$0.1794_{(0.1871)}$&$0.1426_{(0.1044)}$&$0.1093_{(0.0885)}$&$0.0986_{(0.0221)}$&$0.0969_{(0.0277)}$&$0.0953_{(0.0403)}$ \\
			&$\beta_2(\cdot)$& $0.1484_{(0.1252)}$&$0.1034_{(0.0797)}$&$0.0885_{(0.1458)}$&$0.0790_{(0.0444)}$&$0.0696_{(0.0230)}$&$0.0812_{(0.0665)}$ \\
			&$\beta_3(\cdot)$&$0.1801_{(0.2326)}$&$0.1536_{(0.1818)}$&$0.1038_{(0.0169)}$&$0.0576_{(0.0144)}$&$0.0736_{(0.0115)}$&$0.0625_{(0.0405)}$ \\
			&$\delta_1(\cdot)$&$0.2290_{(0.2478)}$&$0.1581_{(0.1418)}$&$0.1458_{(0.0251)}$&$0.1169_{(0.0553)}$&$0.1038_{(0.0263)}$&$0.1200_{(0.1325)}$ \\
			&$\delta_2(\cdot)$&$0.2676_{(0.2928)}$&$0.2166_{(0.2450)}$&$0.15677_{(0.0285)}$&$0.1082_{(0.0331)}$&$0.1071_{(0.0292)}$&$0.1150_{(0.0986)}$ \\\cmidrule(lr){1-8}
			{$0.5$}& $\beta_1(\cdot)$&$0.0586_{(0.0197)}$&$0.0472_{(0.0168)}$&$0.0494_{(0.0175)}$&$0.0464_{(0.0100)}$&$0.0414_{(0.0120)}$&$0.0316_{(0.0111)}$ \\
			&$\beta_2(\cdot)$& $0.0924_{(0.0827)}$&$0.0891_{(0.0562)}$&$0.0657_{(0.0239)}$&$0.0508_{(0.0135)}$&$0.0624_{(0.0179)}$&$0.0439_{(0.0138)}$ \\
			&$\beta_3(\cdot)$&$0.1208_{(0.1334)}$&$0.0953_{(0.0529)}$&$0.0695_{(0.0288)}$&$0.0537_{(0.0146)}$&$0.0541_{(0.0117)}$&$0.0411_{(0.0130)}$ \\
			&$\delta_1(\cdot)$&$0.1382_{(0.0575)}$&$0.1090_{(0.0437)}$&$0.1198_{(0.1086)}$&$0.0798_{(0.0214)}$&$0.0918_{(0.0234)}$&$0.0614_{(0.0113)}$ \\
			&$\delta_2(\cdot)$&$0.1766_{(0.1417)}$&$0.1324_{(0.0612)}$&$0.1280_{(0.1130)}$&$0.0941_{(0.0508)}$&$0.0880_{(0.0150)}$&$0.0703_{(0.0229)}$ \\\cmidrule(lr){1-8}
			{$0.75$}& $\beta_1(\cdot)$&$0.1358_{(0.1649)}$&$0.1339_{(0.0882)}$&$0.0887_{(0.0207)}$&$0.0858_{(0.0153)}$&$0.0833_{(0.0194)}$&$0.0804_{(0.0203)}$ \\
			&$\beta_2(\cdot)$& $0.1435_{(0.0581)}$&$0.1032_{(0.1734)}$&$0.0724_{(0.0250)}$&$0.0661_{(0.0215)}$&$0.0680_{(0.0230)}$&$0.0550_{(0.0202)}$ \\
			&$\beta_3(\cdot)$&$0.1364_{(0.1159)}$&$0.1084_{(0.0771)}$&$0.0650_{(0.0171)}$&$0.0580_{(0.0157)}$&$0.0640_{(0.0134)}$&$0.0518_{(0.0123)}$ \\
			&$\delta_1(\cdot)$&$0.2149_{(0.2819)}$&$0.1352_{(0.0624)}$&$0.1096_{(0.0321)}$&$0.1012_{(0.0505)}$&$0.0951_{(0.0257)}$&$0.0898_{(0.0329)}$ \\
			&$\delta_2(\cdot)$&$0.2388_{(0.3310)}$&$0.1813_{(0.1778)}$&$0.1305_{(0.0344)}$&$0.1196_{(0.0242)}$&$0.1003_{(0.0330)}$&$0.0844_{(0.0282)}$ \\
			\hline
		\end{tabular} 
	} 
	\label{table2:rmise}
\end{table}

\begin{table}[!ht]\tiny
	\centering
	\caption{The mean of Accuracies and standard deviations in subscript brackets  with $\tau=0.25 ,0.5, 0.75$ and error distribution $t(3)$.}
	\vspace{0.2cm}
	\resizebox{\textwidth}{!}{
		\begin{tabular}{ccccccc}
			\hline
			\multirow{2.5}*{$\tau$}&\multicolumn{2}{c}{$n=200$}& \multicolumn{2}{c}{$n=300$}& \multicolumn{2}{c}{$n=400$} \\
			\cmidrule(lr){2-3} \cmidrule(lr){4-5} \cmidrule(lr){6-7}
			& $m=30$ & $m=50$ & $m=30$ & $m=50$ &$ m=30 $& $m=50$   \\ \cmidrule(lr){1-7}
			$0.25$&$0.9612_{(0.0615)}$ & $0.9690_{(0.0560)}$ & $0.9801_{(0.0460)}$ & $0.9798_{(0.0471)}$ & $0.9900_{(0.0296)}$ & $0.9860_{(0.0402)}$ \\ 
			$0.50$& $0.9745_{(0.0503)}$ & $0.9792_{(0.0450)}$ & $0.9846_{(0.0395)}$ & $0.9842_{(0.0422)}$ & $0.9920_{(0.0275)}$ & $0.9898_{(0.0344)}$ \\ 
			$0.75$&$0.9814_{(0.0420)}$ & $0.9850_{(0.0381)}$ & $0.9875_{(0.0356)}$ & $0.9882_{(0.0352)}$ & $0.9936_{(0.0254)}$ & $0.9923_{(0.0294)}$ \\ 
			\hline
		\end{tabular}
	}
	\label{table:acc}
\end{table}

Table \ref{table:acc} lists the mean of Accuracies. It can be seen that the mean of Accuracies tends to 1 as the sample size $n$ or observed data points $m$ increases, which indicates that the proposed method produces the high accuracy of identifying subgroups. The standard deviations are small with light variation in different quantile levels and sample sizes, which further demonstrates good performance of the proposed estimation procedure.

\begin{figure}[!ht]
	\centering
	\includegraphics[height=4cm,width=4.5cm]{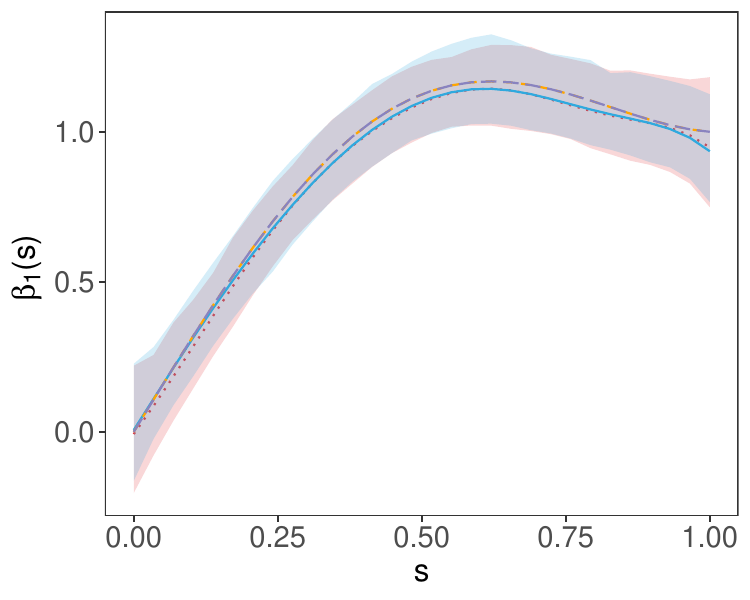}
	\includegraphics[height=4cm,width=4.5cm]{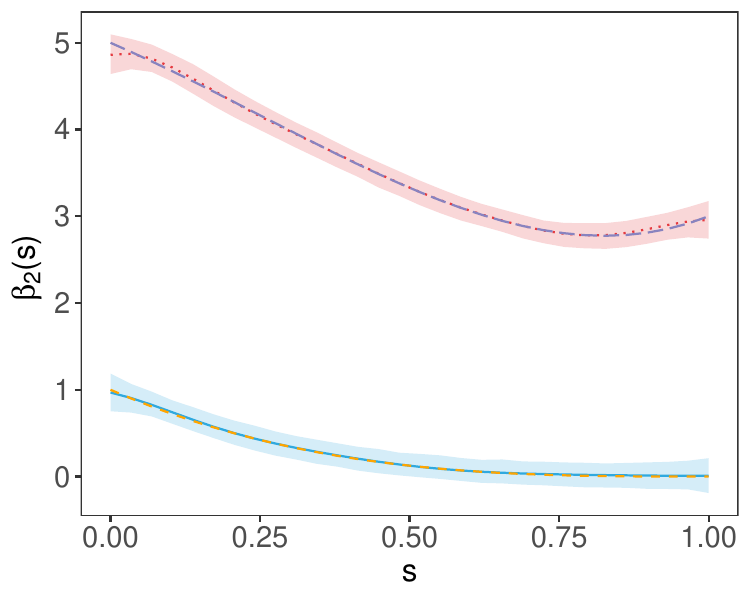}
	\includegraphics[height=4cm,width=4.5cm]{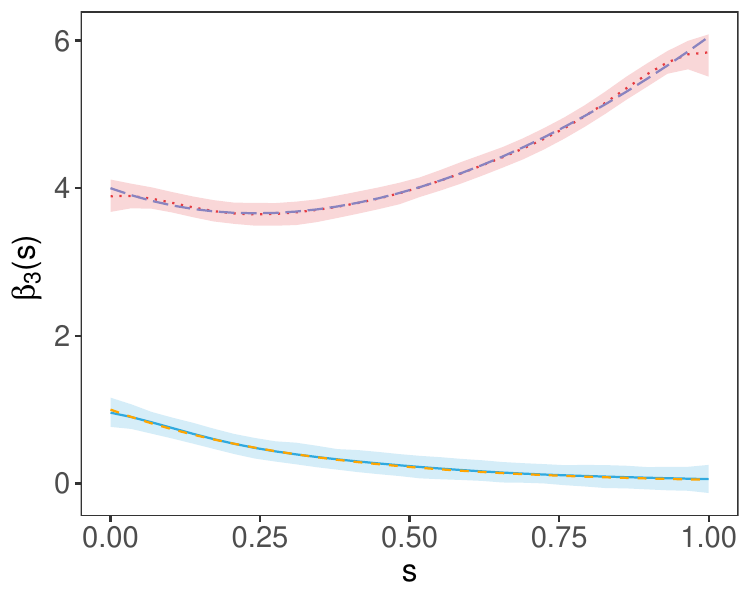}
	\caption{\footnotesize{The true functions (dashed line for Group 0, long dashed line for Group 1) and the mean estimated functions (solid line for Group 0, dotted line for Group 1) from 500 repetitions, along with the 95\% pointwise confidence bands for each component function,  under the condition of subgroup presence at $(\tau, n, m) = (0.75,400,30)$.}}
	\label{fig22.tau75n400m30}
\end{figure}

Once the subgroups are identified, the samples are divided into two subgroups, where Group 0 consists of samples satisfying $ I\left({\bz}_{i}^{T} {\bm \psi}({\tau}) \geq 0  \right)=0$, and Group 1 consists of samples satisfying $ I\left({\bz}_{i}^{T} {\bm \psi}({\tau}) \geq 0 \right)=1$.
Compared to the parameters $(\beta_1(s,\tau),\beta_2(s,\tau), \beta_3(s,\tau))$ in Group 0, the parameters become $(\beta_1(s,\tau),\beta_2(s,\tau)+\theta_1(s,\tau) , \beta_3(s,\tau)+\theta_2(s,\tau))$ in Group 1. Figure \ref{fig22.tau75n400m30} shows the true and the mean estimated functions in each subgroup under the condition of subgroup presence at $(\tau, n, m) = (0.75,400,30)$.
Moreover, Figure \ref{fig22.tau75n400m30} illustrates the disparities in the models between the different subgroups.

\subsection{Performance of hypothesis test} 
In this subsection, we evaluate the performance of the proposed WAST. Specifically, we examine the model outlined in (\ref{model:qr4.1}). The variables $(\bz, \bx, \tilde{\bx})$, the functional coefficients $\bbeta(s,\tau)$, and the error structures are identical to those detailed in Section \ref{sec:estimation}.
Let ${\bm \psi}(\tau) = (1, {\bm \psi}_1(\tau), {\bm \psi}_2(\tau))^{T}$, and we adopt ${\bm \psi}_1(\tau)$ as the negative of the 65\% percentile of $z_{1} + z_{2}{\bm \psi}_2(\tau)$ such that $\bz^{T}{\bm \psi}(\tau)$ divides the population into two subgroups with 35\% and 65\% observations.
We evaluate the power under a sequence of alternative models indexed by $\xi$, that is, $H_1^{\xi}: \bm\theta_{\xi}(s,\tau) = \xi\bm\theta(s,\tau)$, where $\bm\theta(s,\tau)$ is the same as that in Section \ref{sec:estimation}, and we set $\xi \in \{0.0, 0.1, 0.2, 0.3, 0.4, 0.5\}$. It is worthy to note that $\xi=0$ represents the null hypothesis.

\begin{figure}[!ht]
	\centering
	\includegraphics[height=4cm,width=4.5cm]{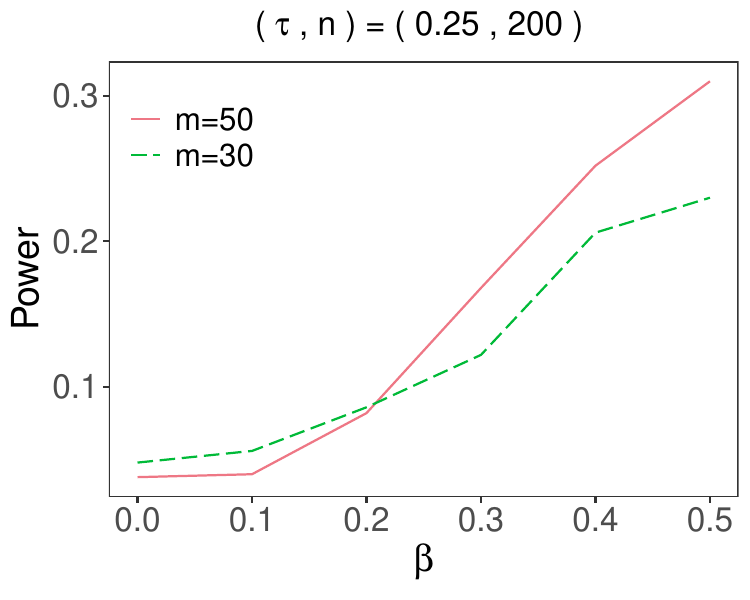}
	\includegraphics[height=4cm,width=4.5cm]{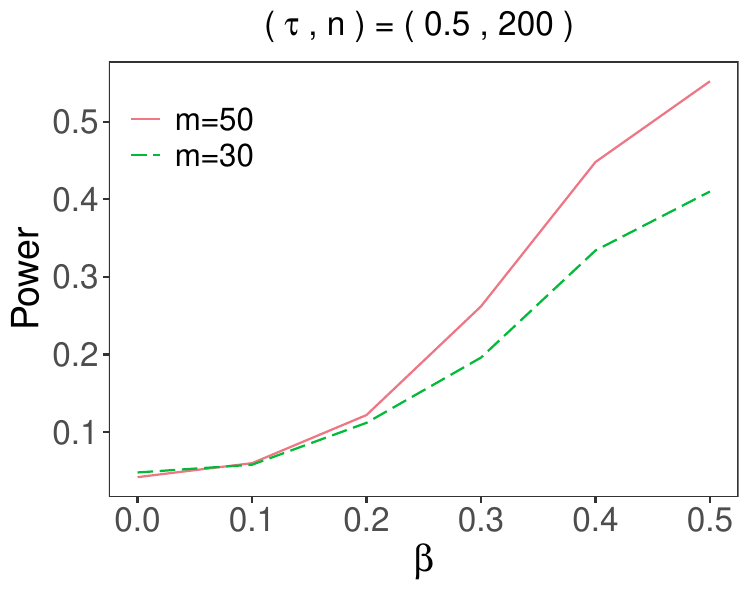}
	\includegraphics[height=4cm,width=4.5cm]{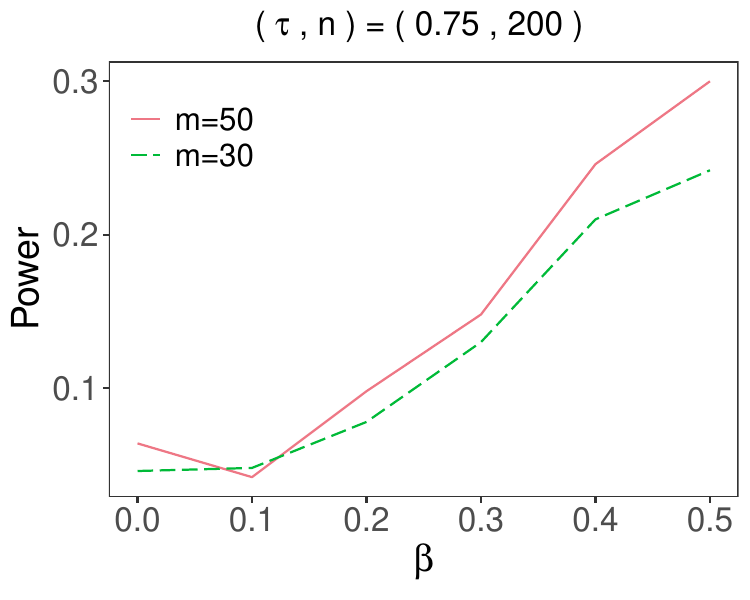}
	\includegraphics[height=4cm,width=4.5cm]{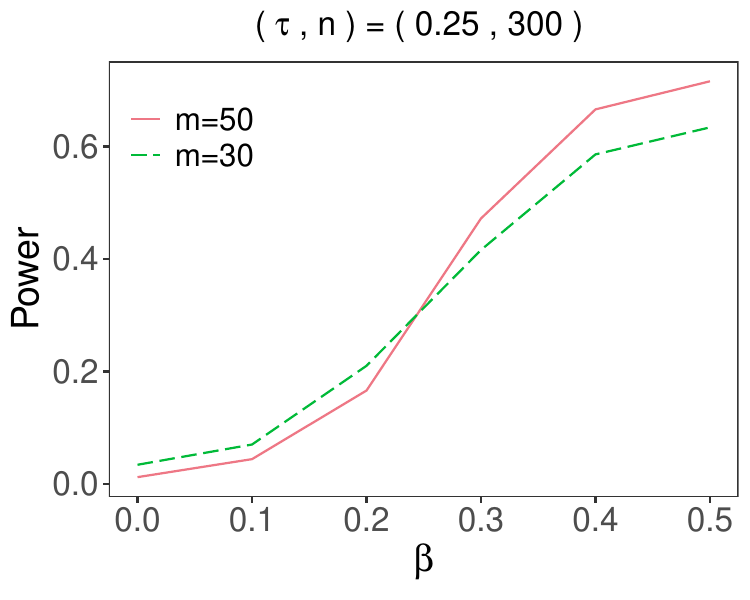}
	\includegraphics[height=4cm,width=4.5cm]{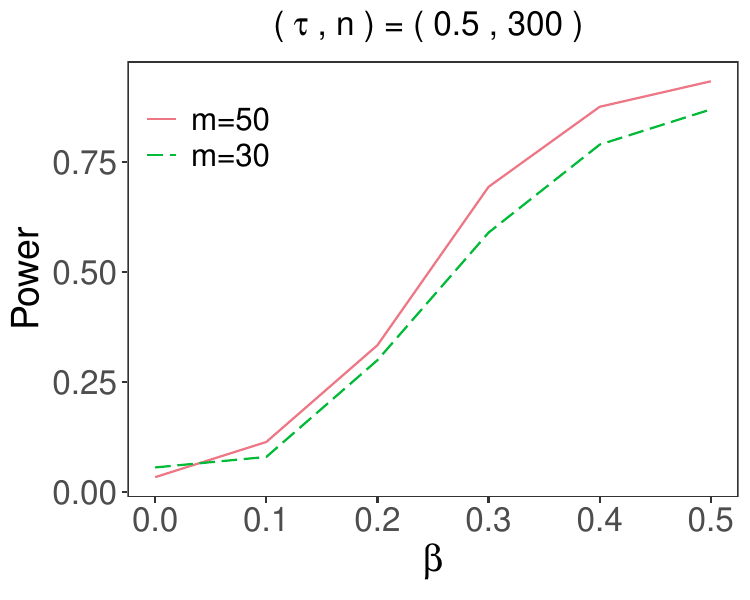}
	\includegraphics[height=4cm,width=4.5cm]{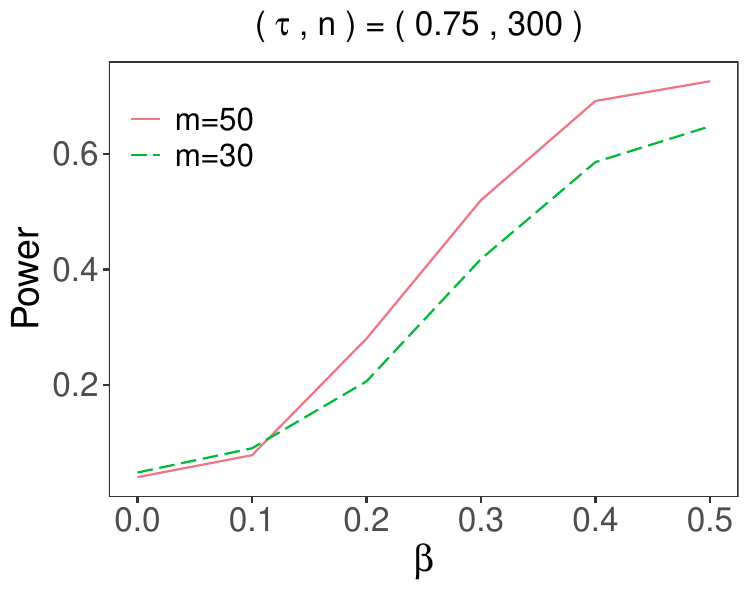}
	
	\includegraphics[height=4cm,width=4.5cm]{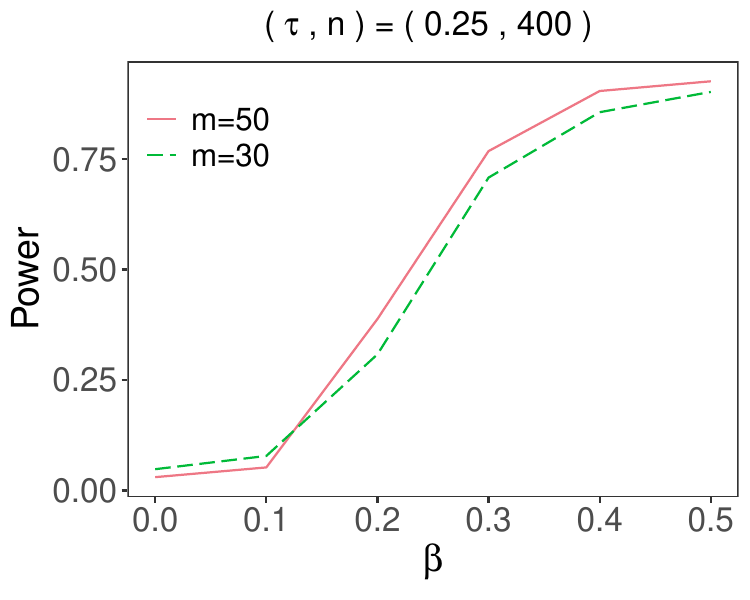}
	\includegraphics[height=4cm,width=4.5cm]{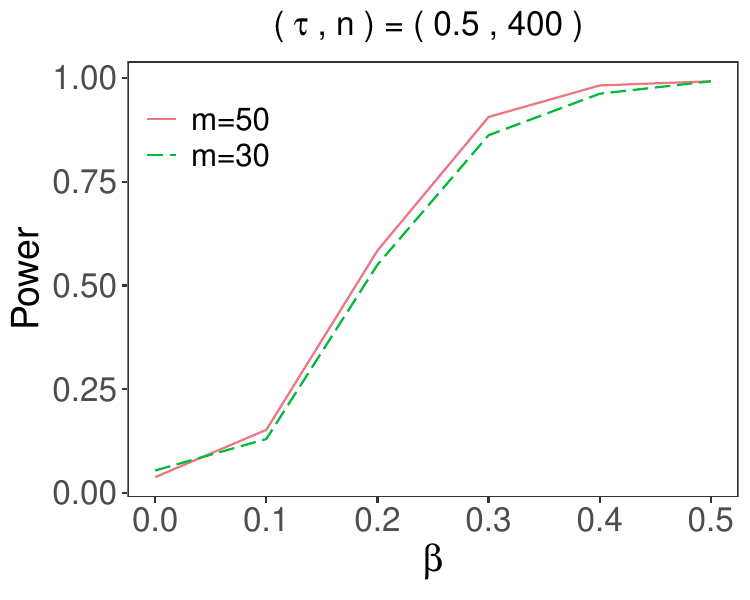}
	\includegraphics[height=4cm,width=4.5cm]{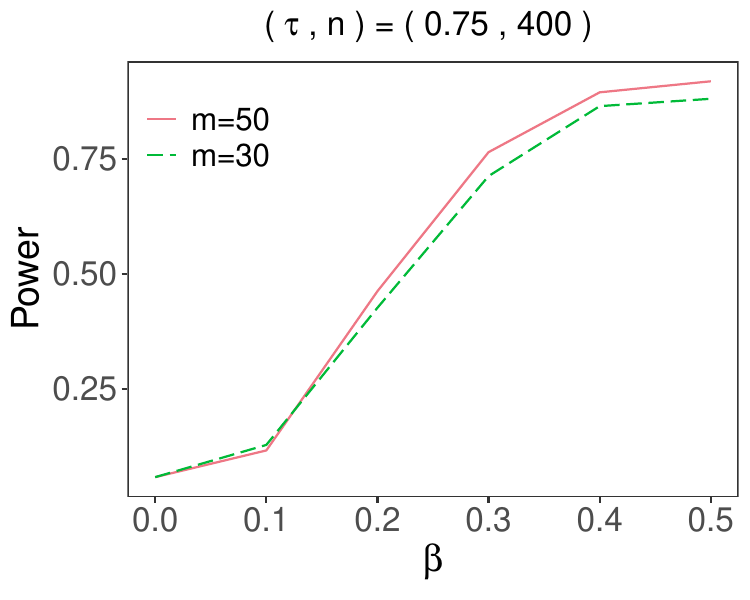}
	
	\caption{\footnotesize{The power curves under $t(3)$  for different observed data points $m=30$ (dashed line) and $m=50$ (solid line) with consistent settings for $n$ and $\tau$. From top to bottom, each row corresponds to $n=200,300,400$, while from left to right, each column corresponds to $\tau=0.25,0.5,0.75$. }}
	\label{fig:powern}
\end{figure}

\begin{figure}[!ht]
	\centering
	\includegraphics[height=4cm,width=4.5cm]{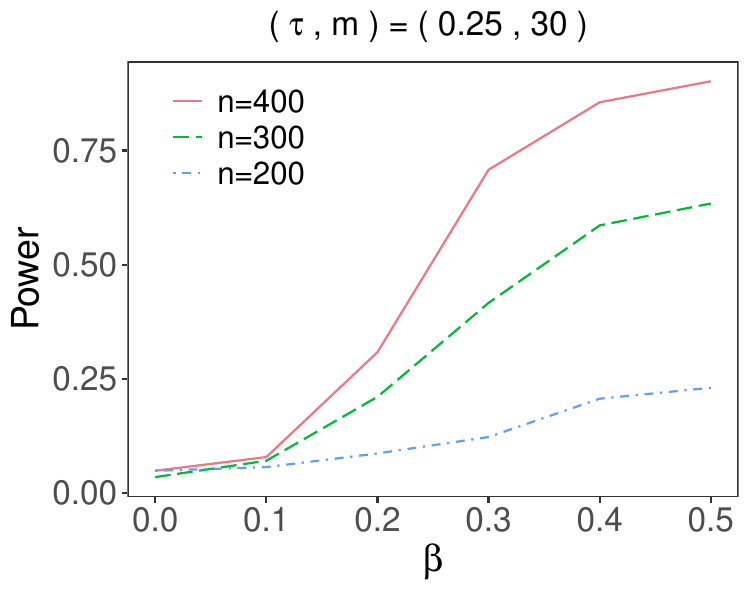}
	\includegraphics[height=4cm,width=4.5cm]{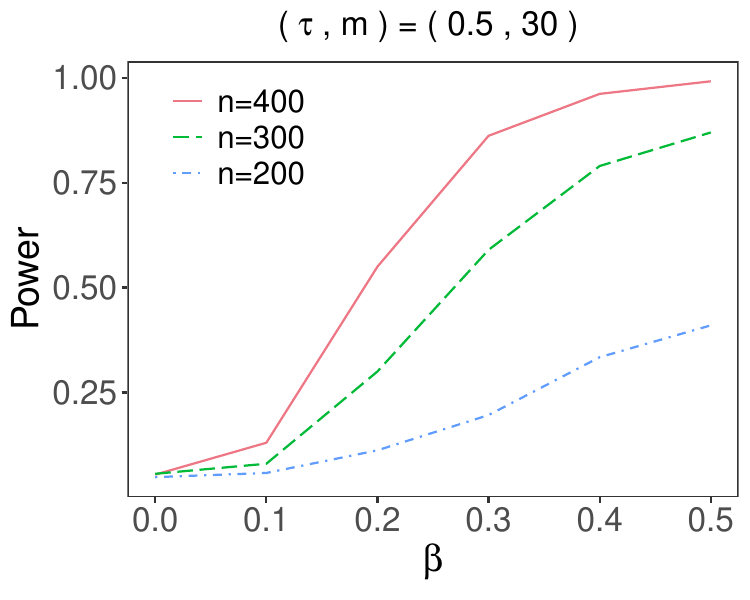}
	\includegraphics[height=4cm,width=4.5cm]{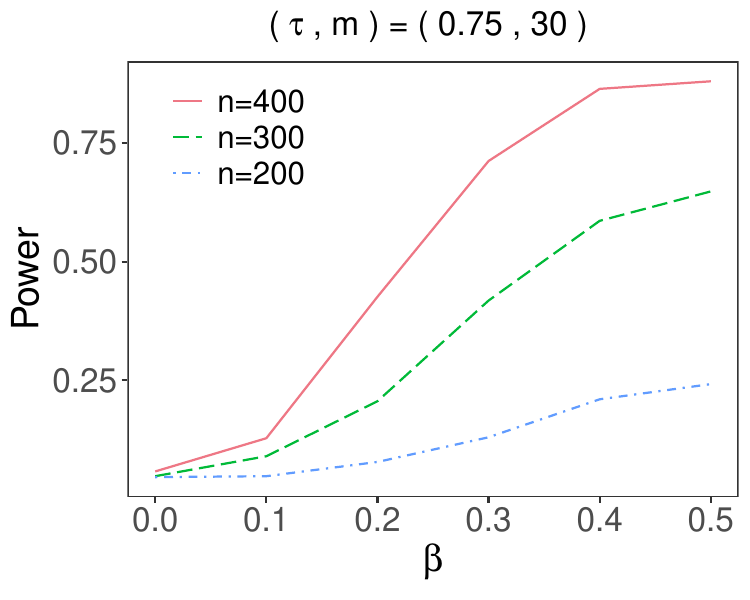}
	\includegraphics[height=4cm,width=4.5cm]{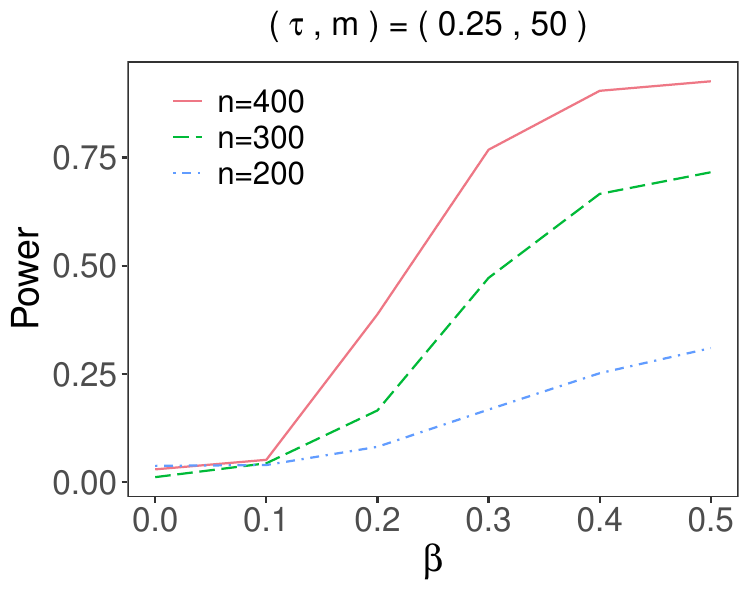}
	\includegraphics[height=4cm,width=4.5cm]{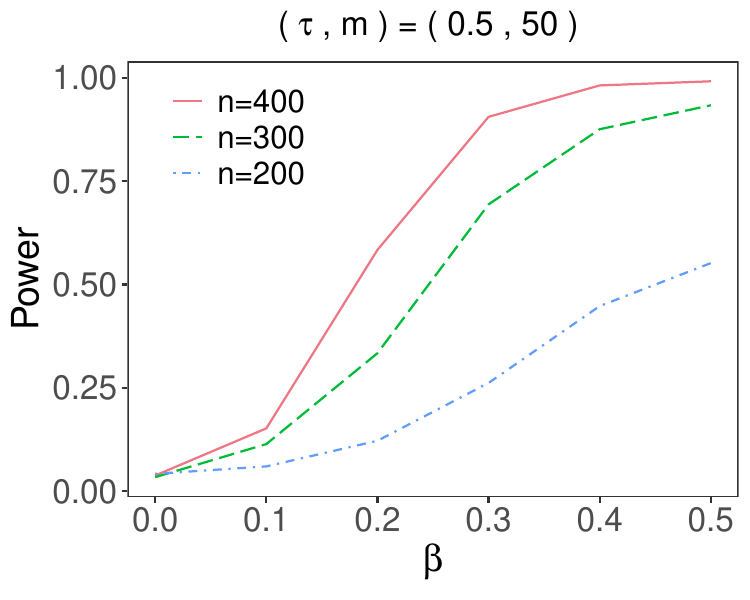}
	\includegraphics[height=4cm,width=4.5cm]{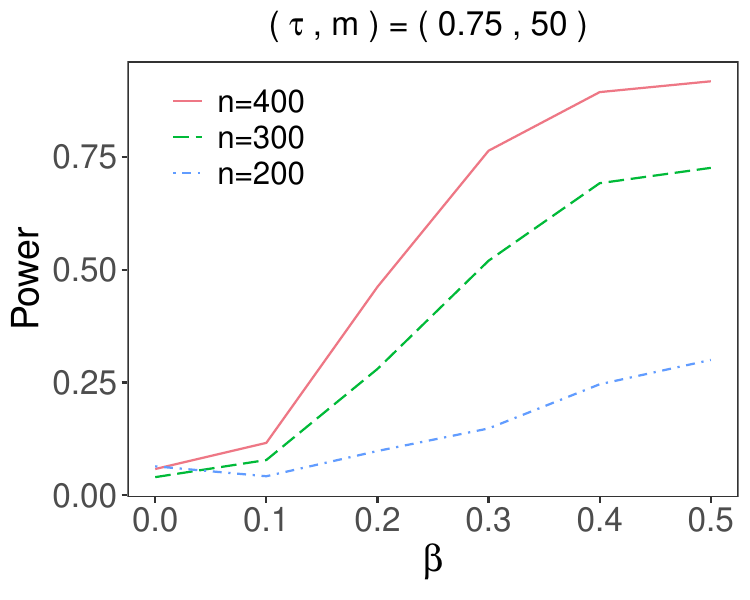}	
	\caption{\footnotesize{The power curves under $t(3)$ for different sample size  $n=200$ (dashdot line), $n=300$ (dashed line) and $n=400$ (solid line) with consistent settings for $m$ and $\tau$. From top to bottom, each row corresponds to $m=30, 50$, while from left to right, each column corresponds to $\tau=0.25,0.5,0.75$.}}
	\label{fig:powerm}
\end{figure}

Both the number of repetitions and the number of bootstraps are set to 500. The sizes ($\xi=0$) and the powers of the WAST test method under the error distribution $t(3)$ are shown in Figure \ref{fig:powern} and \ref{fig:powerm}.
It is observed from Figure \ref{fig:powern} and Figure \ref{fig:powerm} that the sizes in various quantile level $\tau$ and sample size $n$ and $m$ are close to the nominal significance level $0.05$, 
which demonstrates that the proposed WAST can control Type I error well, which verifies the asymptotic results in Theorem \ref{th4}.
Furthermore, the powers increase rapidly to 1  when the alternative hypothesis departs from the null. As expected, the powers for the WAST in Figure \ref{fig:powern} increase as the sample size $n$ grows, and the WAST's powers in Figure \ref{fig:powerm} increase as the observed data points $m$ grows. This verifies the asymptotic results in Theorem \ref{th5}.
To save space here, we report the similar performance of Type I errors and power curves  with errors for the Gaussian and Laplace distributions are presented in Table B5, and Figures B9-B11 of Appendix B in the Supplementary Material.

\section{Case study}\label{sec5}
In this section, we apply the proposed method to learn the subgroups in the Stock dataset and COVID-19 dataset.
To conserve space, the COVID-19 study is provided in the Supplementary Material. 

The Stock dataset contains 182 stocks from the Shanghai A-share market, covering a 59-day period from October 1, 2020 to December 31, 2020. Consider the following functional change-plane quantile regression:
\begin{align*}
      Q_{y_{i}(s)}(\tau | {\bx}_{i}, \widetilde{\bx}_i, {\bz}_{i}, s)={\bx}_{i}^{T} {\bm \beta}(s, \tau)+\widetilde{x}_i{ \theta}(s, \tau) I\left({\bz}_{i}^{T} {\bm \psi}({\tau}) \geq 0\right), 
\end{align*}
where $\{y_i(s), i =1, \cdots, 182\}$ are logrithms of returns of 182 stocks at $m=59$ days,  $\bx = (1, \tilde{x})^T$ with $\tilde{x}$ being the final price at which the stock trades upon the closing of the exchange on the past quarter (lagged price).
The grouping variables  $\boldsymbol{z} = (z_1, 1, z_2, z_3)^T$ consider three financial variables measuring the profitability of a company, including the return on total assets($z_1$), the return on net assets($z_2$), and the operating profit margin($z_3$). The grouping parameters $\bpsi(\tau) = (1, \psi_1(\tau), \psi_2(\tau), \psi_3(\tau))^{T}$, the coefficient functions are ${\bm \beta}(s,\tau) = \left(\beta_0(s,\tau),\beta_1(s,\tau)\right)^T$ and $\theta(s,\tau)$.  All variables are standardized.

We consider the heterogeneous effects of the lagged price on log returns and perform the subgroup testing as
\begin{gather*} 
H_{0}: { \theta}(s,\tau)  \equiv 0, \forall s \in[0,1]\quad \text{ versus } \quad  H_{1}: {\theta}(s,\tau) \neq 0 , \exists s \in[0,1], 
\end{gather*}
with $\tau = 0.25, 0.5, 0.75$, respectively.
By applying the proposed WAST method across the three quantiles, all p-values are less than $10^{-4}$, providing strong evidence for the presence of heterogeneity in stocks.

\begin{figure}[!ht]
	\centering
	\subfigure[]{
	\includegraphics[height=4cm,width=5cm]{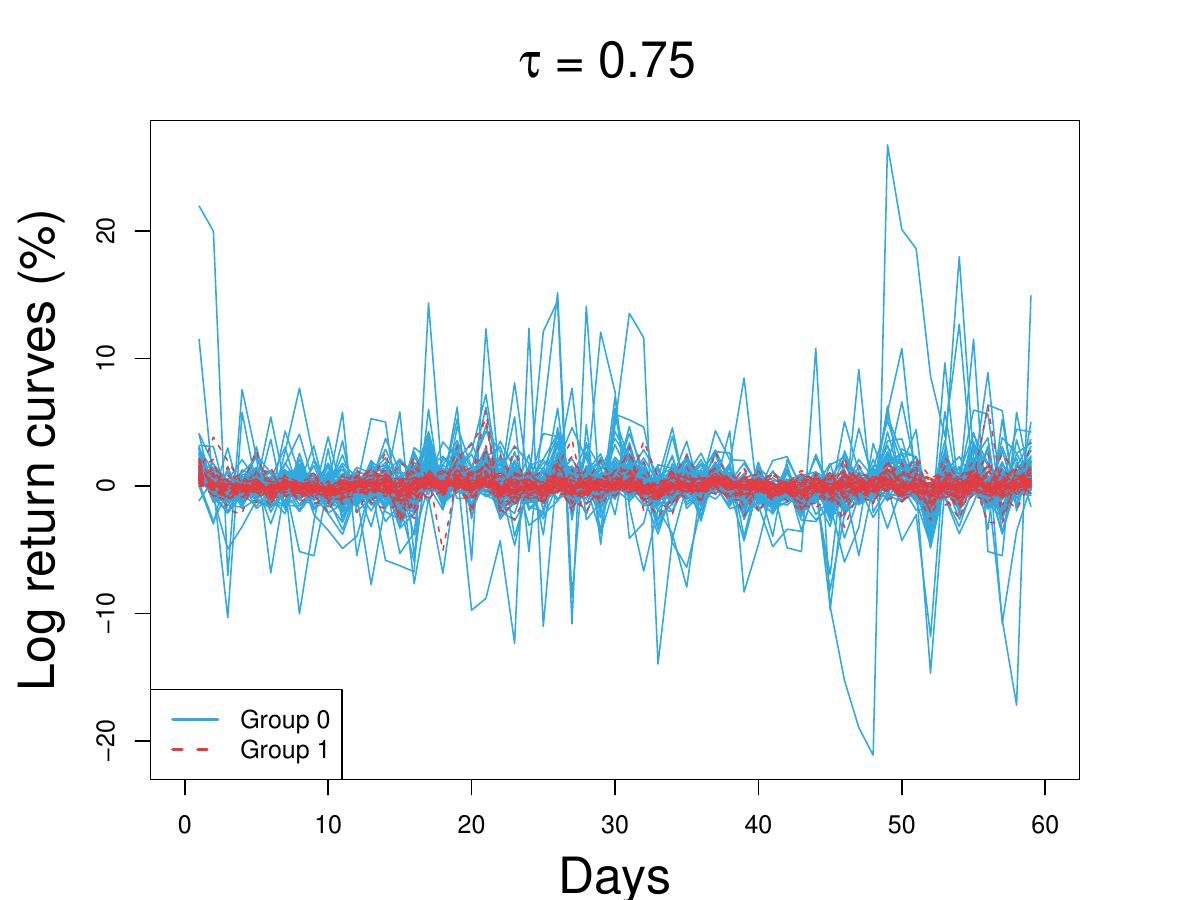}
	}
	\subfigure[]{
    \includegraphics[height=4cm,width=5cm]{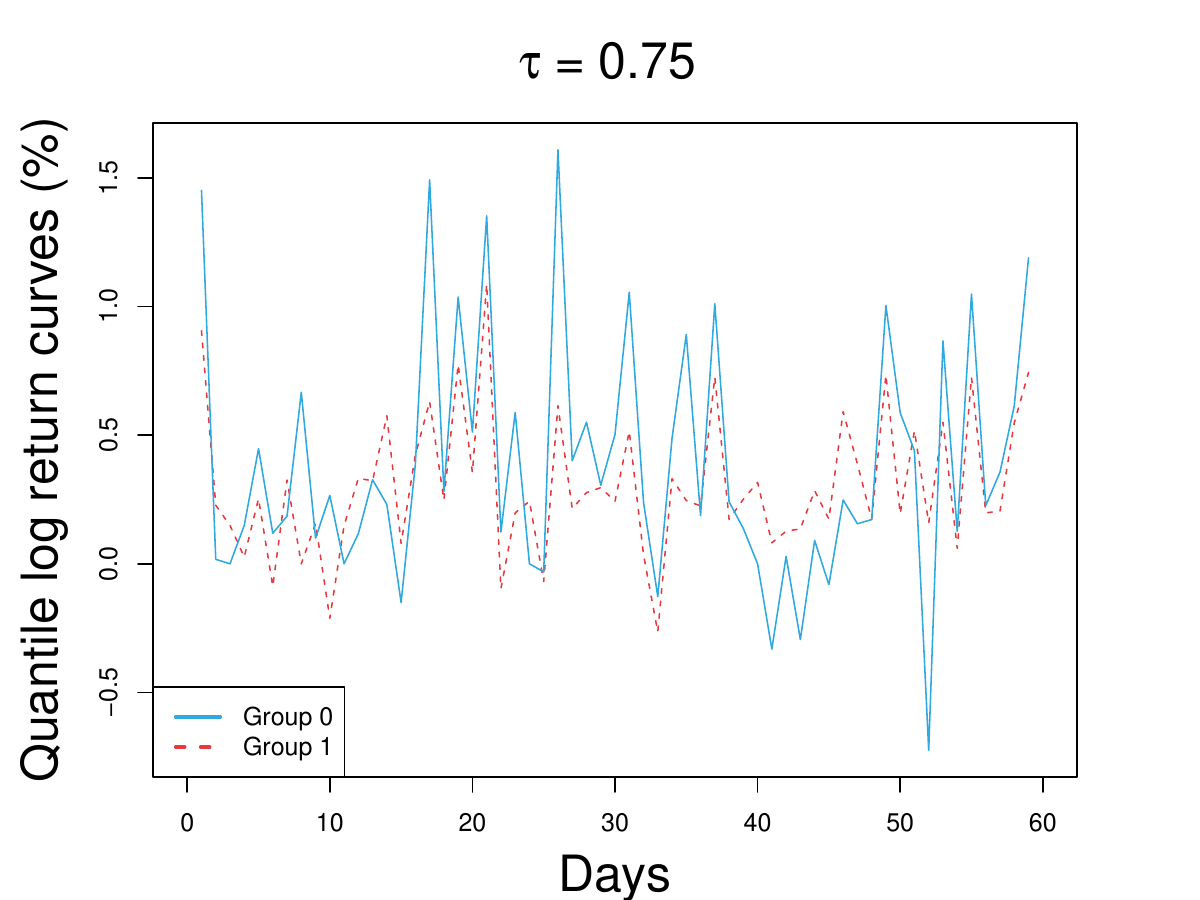}
    }
    \subfigure[]{
	\includegraphics[height=4cm,width=5cm]{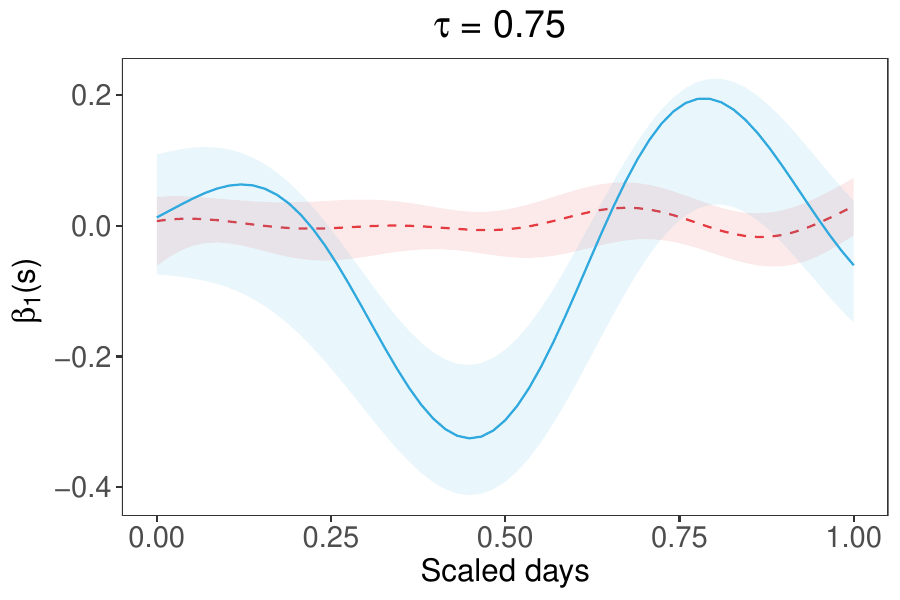}
   }
	\caption{\footnotesize{(a) The log return curves of 182 stocks at the quantile level $\tau = 0.75$; (b) The 0.75-th quantile of the log return curves for the two subgroups, where the dashed line corresponds to Group 1 and the solid line corresponds to Group 0; (c) The estimated function $\hat{\beta}_1(s,0.75)$ on Group 0 (solid line) and Group 1 (dashed line) at 0.75-th quantile level in the stock dataset.}}
	\label{fig:stock_y}
\end{figure}

Take the 0.75-th quantile as an example, the estimator of grouping parameter is $\hat{\bpsi}(0.75) = (1, -0.1884, -0.4581, 0.1470)^T$, which partitions the stocks into two distinct subgroups,  including Group 0 $(\{i: {\bz}_i^T\hat{\bpsi}(0.75) \geq 0 \})$ and Group 1 $(\{i: {\bz}_i^T\hat{\bpsi}(0.75) < 0\})$. Specifically, there are 136 stocks in Group 0 and 36 stocks in Group 1.
Figure \ref{fig:stock_y}(a)-(b) depict the log return curves of the Shanghai A-share stocks in two estimated subgroups. 
Specifically, Figure \ref{fig:stock_y}(a) reveals that
Group 1 is characterized by relatively stable log returns, which suggests more moderate equity market performance, while Group 0 exhibits heightened volatility, indicating more turbulent and unpredictable stock market conditions. Figure \ref{fig:stock_y}(b) depicts the distinct 0.75-th quantile of log return curves for the two subgroups. 
It can be observed that the two subgroups exhibit different levels of volatility, suggesting that the lagged prices of the two subgroups may have distinct impacts on the 0.75-th quantile of log returns. Using a unified model may fail to capture this complexity.
Figure \ref{fig:stock_y}(c) depict the estimated function $\hat{\beta}_1(s,0.75)$ of each subgroup. It can be seen that $\hat{\beta}_1(s,0.75)$ for Group 1 remains close to zero, indicating that the effect of lagged prices on the the 0.75-th quantile of log returns is weak.
The analysis results for the 0.25-th quantile are similar to those for the 0.75-th quantile and are provided in Appendix C of the Supplementary Material.

\section{Conclusion}\label{sec6}
This paper presents a change-plane analysis applied to functional data within the framework of functional response quantile regression, filling the existing gap in subgroup analysis. 
The proposed method can identify and test heterogeneity in non-Gaussian functional responses with scalar predictors, offering great flexibility and robustness.
The RKHS approach enables us to obtain accurate estimators of functional coefficients, while the smoothing method yields standard limiting distributions for the grouping parameters.
Moreover, the asymptotic results demonstrates that there is asymptotic independence between the functional coefficients and the grouping parameters, consistent with findings in change-plane analysis of scalar data.
The proposed ADMM algorithm is computationally flexible and performs well in simulation studies.
For statistic inference, we proposed a novel WAST statistic, which has a closed-form solution by selecting an appropriate weight, thereby reducing the computational burden compared to the methods in \cite{huang2020} and  \cite{ songsui2017}.
The asymptotic properties of the test statistic are established under both null and alternative hypotheses. Extensive simulation studies provide empirical evidence that the test statistic demonstrates strong statistical power.

There are two potential directions for future research. 
One direction is to extend the proposed model to accommodate multiple thresholds, as discussed by \cite{lijialiang2018} and \cite{wang2022}.
We can explore the scenario where multiple thresholds exist and their locations are unknown, which adds flexibility to the model.
Another direction is to apply change-plane analysis to functional linear regression models, where the response is a scalar and the predictors are functional.
This method could be utilized to identify subgroups with enhanced treatment effects in precision medicine. Furthermore, it can be extended to the framework of generalized linear models or the Cox model for survival analysis.
%
%


\begin{center}
	{\large\bf SUPPLEMENTARY MATERIAL}
\end{center}
The Supplementary Material includes additional simulation results, empirical analyses, and the proofs for theorems in the paper.

\bibliographystyle{agsm} 
\bibliography{qcp}

@article{wahba1990,
	title={Spline models for observational data},
	author={ Wahba, G. },
	journal={watson research center},
	year={1990}
}

@article{wangxiao2020,
	title={High-Dimensional spatial quantile function-on-scalar Regression},
	author={Zhengwu Zhang and Xiao Wang and Linglong Kong and Hongtu Zhu},
	journal={Journal of the American Statistical Association},
	year={2022},
	volume={117},
	pages={1563-1578}
}

@article{Boyd2011,
	title={Distributed optimization and statistical learning via the alternating direction method of multipliers},
	author={Stephen P. Boyd and Neal Parikh and Eric King-Wah Chu and Borja Peleato and Jonathan Eckstein},
	journal={Now Foundations and Trends},
	year={2011},
	volume={3},
	pages={1-122}
}

@article{songsui2017,
	title={Change-Plane analysis for subgroup detection and sample size calculation},
	author={Ailin Fan and Rui Song and Wenbin Lu},
	journal={Journal of the American Statistical Association},
	year={2017},
	volume={112},
	pages={769-778}
}

@article{lijialiang2018,
	title={Multi-threshold change plane model: estimation theory and applications in subgroup identification},
	author={ Li, Jialiang  and  Li, Yaguang  and  Jin, Baisuo },
	journal={Statistics in Medicine},
	year={2021},
	volume={40},
	Pages={3440-3459}
}

@article{seolinton2005,
	title={A smoothed least squares estimator for threshold Regression Models},
	author={Myung Hwan Seo and Oliver B. Linton},
	journal={Journal of Econometrics},
	year={2007},
	volume={141},
	pages={704-735}
}

@article{zhangyy2021,
	title={Single-index thresholding in quantile regression},
	author={Yingying Zhang and Huixia Judy Wang and Zhongyi Zhu},
	journal={Journal of the American Statistical Association},
	year={2021},
	volume={141},
	pages={1-42}
}

@article{wang2016,
	title={Review of functional data analysis},
	author={Jane Wang and Jeng Min Chiou and Hans Georg M\"{u}eller},
	journal={Annual Review of Statistics and Its Application},
	year={2016},
	volume={3},
	pages={257-295}
}

@book{ramsay2005,
	title={Functional data analysis},
	author={J. O. Ramsay  and B. W. Silverman },
	year={2005},
	publisher={Springer}
}

@article{deng2022,
	title={Maximum likelihood estimation for Cox proportional hazards model with a change hyperplane},
	author={Yu Deng and Jianwen Cai and Donglin Zeng},
	journal={Statistica Sinica},
	year={2022},
	volume={32},
	pages={983-1006}
}

@article{huang2020,
	title={Threshold-based subgroup testing in logistic regression models in two‐phase sampling designs},
	author={Ying Huang and Juhee Cho and Youyi Fong},
	journal={Journal of the Royal Statistical Society: Series C (Applied Statistics)},
	year={2021},
	volume={70},
	pages={291-311}
}

@article{jiang2021,
	title={Cluster analysis with regression of non‐Gaussian functional data on covariates},
	author={Jiakun Jiang and Huazhen Lin and Heng Peng and Gang-Zhi Fan and Yi Li},
	journal={Canadian Journal of Statistics},
	year={2021},
	volume={50},
	pages={221-240}
}

@article{yao2017,
	title={Regularized partially functional quantile regression},
	author={Fang Yao and Shivon Sue-Chee and Fan Wang},
	journal={Journal of Multivariate Analysis},
	year={2017},
	volume={156},
	pages={39-56}
}

@article{zhou2023,
	title={Functional response quantile regression Model.},
	author={Xingcai Zhou and Dehan Kong and A. B. Kashlak and Linglong Kong and R. Karunamuni and Hongtu Zhu},
	journal={Statistica Sinica},
	year={2023},
	volume={33},
	pages={2643-2667}
}

@article{wei2023,
	title={Subgroup analysis for longitudinal data based on a partial linear varying coefficient model with a change plane},
	author={Kecheng Wei and Guoyou Qin and Zhongyi Zhu},
	journal={Statistics in Medicine},
	year={2023},
	volume={42},
	pages={3716-3731}
}

@article{shen2015,
	title={Inference for subgroup analysis with a structured logistic-normal mixture model},
	author={Juan Shen and Xuming He},
	journal={Journal of the American Statistical Association},
	year={2015},
	volume={110},
	pages={303-312}
}

@article{ma2017,
	title={A concave pairwise fusion approach to subgroup analysis},
	author={Shujie Ma and Jian Huang},
	journal={Journal of the American Statistical Association},
	year={2017},
	volume={112},
	pages={410-423}
}

@article{liu2024,
	title={Efficient subgroup testing with large numbers of grouping variables},
	author={Xu Liu and Jian Huang and Yong Zhou and Feipeng Zhang and Panpan Ren},
	year={2024},
	journal={arxiv.org/abs/2408.00602},
}

@article{wang2021,
	title={Latent group detection in functional partially linear regression models},
	author={Wu Wang and Ying Sun and Huixia Judy Wang},
	journal={Biometrics},
	year={2021},
	volume={79},
	pages={280-291}
}

@article{yuan2011,
	title={A reproducing kernel Hilbert space approach to functional linear regression},
	author={Ming Yuan and T. Tony Cai},
	journal={The Annals of Statistics},
	year={2010},
	volume={38},
	pages={3412-3444}
}

@article{shang2015,
	title={Nonparametric inference in generalized functional linear models},
	author={Zuofeng Shang and Guang Cheng},
	journal={The Annals of Statistics},
	year={2015},
	volume={43},
	pages={1742-1773}
}

@article{li2007,
	title={Quantile regression in reproducing kernel Hilbert spaces},
	author={Youjuan Li and Yufeng Liu and Ji Zhu},
	journal={Journal of the American Statistical Association},
	year={2007},
	volume={102},
	pages={255-268}
}

@article{zhu2012,
	title={Multivariate varying coefficient model for functional responses},
	author={Hong-Tu Zhu and Runze Li and Linglong Kong},
	journal={The Annals of Statistics},
	year={2012},
	volume={40},
	pages={2634-2666}
}

@article{li2007on,
	title={On rates of convergence in functional linear regression},
	author={Yehua Li and Tailen Hsing},
	journal={Journal of Multivariate Analysis},
	year={2007},
	volume={98},
	pages={1782-1804}
}

@article{bai1998,
	title={Estimating and testing linear models with multiple structural changes},
	author={Bai, J. and P. Perron},
	journal={Econometrica},
	year={1998},
	volume={66},
	pages={47-78}
}

@article{minh2016,
	title={A Unifying Framework in Vector-valued Reproducing Kernel Hilbert Spaces for Manifold Regularization and Co-Regularized Multi-view Learning},
	author={H\'{a} Quang Minh  and  Loris Bazzani  and  Vittorio Murino},
	journal={Journal of Machine Learning Research},
	year={2016},
	number={17},
	pages={1-72}
}

@article{cai2011aos,
	title={Optimal estimation of the mean function based on discretely sampled functional data: Phase transition},
	author={T. Tony Cai and Ming Yuan},
	journal={The Annals of Statistics},
	year={2011},
	volume={39},
	pages={2330-2355}
}

@article{wang2022,
	title={Multi-Threshold structural equation model},
	author={Jingli Wang and Jialiang Li},
	journal={Journal of Business and Economic Statistics},
	year={2022},
	volume={41},
	pages={377-387}
}

@article{hansen2011,
	title={Threshold autoregression in economics},
	author={Bruce E. Hansen},
	journal={Statistics and Its Interface},
	year={2011},
	volume={4},
	pages={123-127}
}

@article{li2021,
	title={Clusterwise functional linear regression models},
	author={Ting Li and Xinyuan Song and Yingying Zhang and Hongtu Zhu and Zhongyi Zhu},
	journal={Computational Statistics and Data Analysis},
	year={2021},
	volume={158},
	pages={107192}
}

@article{zhang2022,
	title={Subgroup analysis for high-dimensional functional regression},
	author={Xiaochen Zhang and Qingzhao Zhang and Shuangge Ma and Kuangnan Fang},
	journal={Journal of Multivariate Analysis },
	year={2022},
	volume={192},
	pages={105100}
}

@article{sun2024,
	title={Subgroup analysis for the functional linear model},
	author={Yifan Sun and Ziyi Liu and Wu Wang},
	journal={Journal of Statistical Planning and Inference},
	year={2024},
	volume={231},
	pages={106120}
}

@article{yao2011,
	title={Functional mixture regression},
	author={Fang Yao and Yuejiao Fu and Thomas C. M. Lee},
	journal={Biostatistics},
	year={2011},
	volume={12},
	pages={341-353}
}

@article{yu2020,
	title={Threshold regression with a threshold boundary},
	author={Ping Yu and Xiaodong Fan},
	journal={Journal of Business and Economic Statistics},
	year={2021},
	volume={39},
	pages={953-971}
}

@article{su2019,
	title={Common threshold in quantile regressions with an application to pricing for reputation},
	author={Liangjun Su and Pai Xu},
	journal={Econometric Reviews},
	year={2019},
	volume={38},
	pages={417-450}
}

@article{su2020,
	title={Testing and estimation of social network dependence with time to event data},
	author={Lin Su and Wenbin Lu and Rui Song and Danyang Huang},
	journal={Journal of the American Statistical Association},
	year={2020},
	volume={115},
	pages={570-582}
}

@article{kang2022,
	title={Inference for change-plane regression},
	author={Chaeryon Kang and Hunyong Cho and Rui Song and Moulinath Banerjee and Eric B. Laber and Michael R. Kosorok},
	year={2022},
	journal={arXiv:2206.06140}
}

\end{document}